\documentclass[a4paper,11pt]{article}
\usepackage[inline]{enumitem}
\usepackage[multidot]{grffile} 
\usepackage{dcolumn}
\usepackage{bm}
\usepackage{amsmath}
\usepackage{amsthm}
\usepackage{framed}
\usepackage{amsfonts}
\usepackage{bbold}
\usepackage{dsfont}
\usepackage{amssymb}
\usepackage{color}
\usepackage{latexsym}
\usepackage{stackrel}
\usepackage{slashed} 
\usepackage{empheq}
\usepackage{fancybox}
\usepackage{pstricks}
\usepackage{indentfirst}
\usepackage{mathrsfs}
\usepackage{tablefootnote}
\usepackage{longtable}
\usepackage{multirow}
\usepackage{epsfig,psfrag}
\usepackage{subfigure}
\usepackage{mathtools}
\usepackage{setspace} 
\usepackage[utf8]{inputenc} 
\usepackage[scientific-notation=true]{siunitx} 
\usepackage{verbatim}
\usepackage{jheppub}
\usepackage[T1]{fontenc} 
\graphicspath{{fig/}}
\newcommand{\uvec}{\boldsymbol}
\newcommand{\hvec}[1]{\hat{\boldsymbol #1} }
\newcommand{\ud}{\text{d}}
\definecolor{Magenta}{rgb}{0.75,0.05,0.75}

\definecolor{LYellow}{rgb}{1., 0.894118, 0.584314}

\newtheorem*{conjecture}{Conjecture}
\theoremstyle{conjecture}

\title{Nucleon relativistic weak-neutral axial-vector four-current distributions}
\author[a,*]{Yi Chen}
\affiliation[a]{Interdisciplinary Center for Theoretical Study and Department of Modern Physics, University of Science and Technology of China, Hefei, Anhui 230026, China}
\emailAdd{physchen@mail.ustc.edu.cn}
\note[*]{Corresponding author}
\abstract{Relativistic full weak-neutral axial-vector four-current distributions inside a general spin-$\frac{1}{2}$ hadron are systematically studied for the first time, where the second-class current contribution associated with the induced pseudotensor form factor (FF) is included. We clearly demonstrate that the 3D axial charge distribution, being parity-odd in the Breit frame, is in fact related to the induced pseudotensor FF $G_T^Z(Q^2)$ rather than the axial FF $G_A^Z(Q^2)$. We study the frame-dependence of full axial-vector four-current distributions for a moving hadron, and compare them with their light-front counterparts. We revisit the role played by the Melosh rotation, and understand more easily and intuitively the origins of distortions appearing in light-front distributions (relative to the Breit frame ones) using the conjecture that we propose in this work. In particular, we show that the second-class current contribution, although explicitly included, does not contribute in fact to the mean-square axial and spin radii. We finally illustrate our results in the case of a proton using the weak-neutral axial-vector FFs extracted from experimental data.}

\begin{document} 
\maketitle

\section{Introduction}
\label{Introduction}

Nucleons (i.e. protons and neutrons) are key hadrons to study for understanding quantum chromodynamics (QCD), since they are responsible for more than $99\%$ of the visible-matter mass in the universe~\cite{Gao:2021sml}. Protons, in particular, also hold another unique role of being the only stable composite building blocks in nature~\cite{Li:2022sqg}. Due to the complicated nonperturbative dynamics of their quark and gluon degrees of freedom, nucleons inherit particularly rich and intricate internal structures. When it comes to the study of properties and internal structures of the nucleon in the weak sector via elastic or quasielastic (anti)neutrino-(anti)nucleon scatterings~\cite{Singh:1971md,Mann:1973pr,Barish:1977qk,Baker:1981su,Miller:1982qi,Kitagaki:1983px,Horstkotte:1981ne,Ahrens:1986xe,Allasia:1990uy,Kitagaki:1990vs,K2K:2006odf,MiniBooNE:2010bsu,MiniBooNE:2010xqw,CLAS:2012ich,MINERvA:2013bcy,MiniBooNE:2013dds,Meyer:2016oeg,MINERvA:2023avz,Pate:2008va,Pate:2024acz}, nucleon axial-vector form factors (FFs) become particularly important.

Axial-vector FFs are Lorentz-invariant functions that describe how the hadron reacts with the incoming (anti)neutrino in a scattering reaction, encoding therefore very clean internal axial charge and spin information of the hadron in the weak sector since (anti)neutrinos participate only in weak interactions. In the Standard Model, there are in general two types of axial-vector FFs of a hadron in the weak sector: the weak-charged ones via the weak-charged current interactions mediated by the $W^\pm$ bosons, and the weak-neutral ones via the weak-neutral current interactions mediated by the $Z^0$ bosons. These axial-vector FFs also serve as important quantities for constraining the systematic uncertainties of high-precision measurements in (anti)neutrino oscillation experiments~\cite{OPERA:2019kzo,DoubleChooz:2019qbj,T2K:2024wfn,DUNE:2024wvj,KamLAND:2022ptk,DayaBay:2024nip,IceCube:2024xjj,JUNO:2024jaw,NOvA:2024imi}. On the theory side, tremendous progress has been reported in the last few years from the first-principle lattice QCD side~\cite{Bhattacharya:2016zcn,Liang:2016fgy,Green:2017keo,Gupta:2017dwj,Yao:2017fym,Hasan:2017wwt,Ishikawa:2018rew,Shintani:2018ozy,Hasan:2019noy,Jang:2019vkm,RQCD:2019jai,Lin:2020rxa,Alexandrou:2020okk,Park:2021ypf,Alexandrou:2021wzv,Ishikawa:2021eut,Meyer:2021vfq,Schulz:2021kwz,Alexandrou:2021jok,Lin:2021brq,Alexandrou:2022yrr,Djukanovic:2022wru,Lin:2022nnj,Koponen:2022gfe,Jang:2023zts,Alexandrou:2023qbg}; theoretical evaluations of the nucleon axial-vector FFs and their contributions to the associated cross sections based on chiral perturbation theory and various models/approaches are still rapidly developing~\cite{Meissner:1986xg,Meissner:1986js,Bernard:1994wn,Ohlsson:1998bk,Barquilla-Cano:2002btm,Silva:2005fa,Schindler:2006jq,Aliev:2007qu,Sharma:2009hg,Eichmann:2011pv,Liu:2014owa,Dahiya:2014jfa,Ramalho:2015jem,Anikin:2016teg,Mamedov:2016ype,Hashamipour:2019pgy,Mondal:2019jdg,Zhang:2019iyx,Jun:2020lfx,Chen:2020wuq,Chen:2021guo,Ahmady:2021qed,Sauerwein:2021jxb,Xu:2021wwj,Atayev:2022omk,Chen:2022odn,Cheng:2022jxe,Liu:2022ekr,Irani:2023lol,Kaur:2023lun,Ramalho:2024tdi,Tomalak:2022xup,Sobczyk:2024ecl,SajjadAthar:2024etq}. For (recent) reviews on nucleon axial-vector FFs and associated physics of (anti)neutrino interactions, see e.g. Refs.~\cite{LlewellynSmith:1971uhs,Gourdin:1974iq,Bernard:1995dp,Bernard:2001rs,Gorringe:2002xx,Beise:2004py,Gallagher:2011zza,Formaggio:2012cpf,Morfin:2012kn,Gonzalez-Jimenez:2011qkf,Alvarez-Ruso:2014bla,Mosel:2016cwa,Krebs:2016rqz,NuSTEC:2017hzk,Hill:2017wgb,Meyer:2022mix,SajjadAthar:2022pjt}.

According to textbooks, charge distributions can be defined in the Breit frame (BF) in terms of three-dimensional (3D) Fourier transform of the Sachs electric FF~\cite{Ernst:1960zza,Sachs:1962zzc,Gao:2021sml}. However, relativistic recoil corrections spoil their interpretation as probabilistic distributions~\cite{Yennie:1957rmp,Breit:1964ga,Kelly:2002if,Burkardt:2000za,Belitsky:2003nz,Jaffe:2020ebz}. In position space, a probabilistic density interpretation is tied to Galilean symmetry that implies the invariance of inertia under the change of frames. In a relativistic theory, however, inertia becomes a frame-dependent concept because of Lorentz symmetry. The only way out is to switch to the light-front (LF) formalism~\cite{Brodsky:1997de} where a Galilean subgroup of the Lorentz group is singled out~\cite{Susskind:1967rg,Kogut:1969xa}, allowing therefore a nice probabilistic interpretation~\cite{Burkardt:2002hr,Miller:2007uy,Carlson:2007xd,Alexandrou:2008bn,Alexandrou:2009hs,Carlson:2009ovh,Gorchtein:2009qq,Miller:2010nz,Miller:2018ybm,Freese:2023jcp,Freese:2023abr}. The price to pay is that besides losing the longitudinal spatial dimension\footnote{We note that Miller and Brodsky~\cite{Miller:2019ysh} recently demonstrated at the wavefunction level that frame-independent and three-dimensional LF coordinate-space wavefunctions can be obtained by using the dimensionless, frame-independent longitudinal coordinate $\tilde z$.}, these LF distributions also exhibit various distortions owing to the particular LF perspective, Lorentz effects and complicated Wigner-Melosh rotations~\cite{Jacob:1959at,Durand:1962zza,Melosh:1974cu,Lorce:2011zta,Polyzou:2012ut,Li:2015hew}, which are sometimes hard to reconcile with an intuitive picture of the system in 3D at rest.

The quantum phase-space formalism distinguishes itself by the fact that the requirement of a strict probabilistic interpretation is relaxed and replaced by a milder quasiprobabilistic picture~\cite{Wigner:1932eb,Hillery:1983ms,Bialynicki-Birula:1991jwl}. This approach is quite appealing since it allows one to define in a consistent way relativistic spatial distributions inside a target with arbitrary spin and arbitrary average momentum~\cite{Lorce:2017wkb,Lorce:2018zpf,Lorce:2018egm,Lorce:2020onh,Lorce:2021gxs,Lorce:2022jyi,Lorce:2022cle,Chen:2022smg,Chen:2023dxp,Chen:2024oxx,Lorce:2024ipy,Lorce:2024ewq,Kim:2021kum,Kim:2022bia,Hong:2023tkv}, providing a smooth and natural connection between the BF and essentially the LF pictures\footnote{Strictly speaking, the smooth connection is between the BF and infinite-momentum frame (IMF) distributions, see e.g., Refs.~\cite{Lorce:2020onh,Chen:2022smg,Chen:2023dxp,Chen:2024oxx,Lorce:2018egm,Lorce:2022jyi,Lorce:2024ipy,Lorce:2024ewq,Kim:2021kum,Kim:2022bia,Hong:2023tkv}. However, IMF distributions coincide most of the time with the corresponding LF distributions~\cite{Chen:2023dxp}.}, and allowing one to explicitly trace the spatial distortions caused by Lorentz boosts and Wigner rotations for any spin-$j$ hadrons under the protection of Poincar\'e symmetry.

As an extension of our recent work~\cite{Chen:2024oxx}, we study in this work in detail the relativistic full weak-neutral axial-vector four-current distributions inside a general spin-$\tfrac{1}{2}$ hadron (e.g., the proton), where the second-class current contribution associated with the induced pseudotensor FF is newly taken into account. We explicitly demonstrate that the relativistic 3D weak-neutral axial charge distribution, being parity-odd in the BF, is in fact related to the weak-neutral induced pseudotensor FF $G_T^Z(Q^2)$ rather than the axial FF $G_A^Z(Q^2)$. This clarifies and reconfirms that the quantity $R_A^2$, see the Eq.~(1) of Ref.~\cite{Chen:2024oxx}, is evidently not the 3D mean-square axial radius of a spin-$\frac{1}{2}$ target.

We also study the full weak-neutral axial-vector four-current distributions for a moving hadron focusing on their frame-dependence, and compare them with their LF counterparts. In particular, we explicitly reproduce the LF axial-vector four-current amplitudes via the proper IMF limit of the corresponding elastic frame (EF) amplitudes, which together with our recent works~\cite{Chen:2022smg,Chen:2023dxp,Chen:2024oxx,Lorce:2024ipy,Lorce:2024ewq} inspires us to propose the following conjecture: {\em Any light-front amplitudes for well-defined light-front distributions in principle can be explicitly reproduced from the corresponding elastic frame amplitudes in the proper infinite-momentum frame limit}. As a reward, we can understand more easily and intuitively the origins of distortions appearing in LF distributions (relative to the BF ones). For completeness, both 3D and 2D transverse mean-square axial and spin radii in different frames are rederived. We show in particular that the second-class current contribution, although explicitly included in our calculations, does not contribute in fact to the mean-square axial and spin radii.

The paper is organized as follows. In Sec.~\ref{sec:Weak-neutral axial-vector form factors}, we present the full matrix elements of the weak-neutral axial-vector four-current operator and associated weak-neutral axial-vector FFs for a general spin-$\frac{1}{2}$ hadron. The key ingredients of the quantum phase-space formalism are briefly given in Sec.~\ref{sec:The quantum phase-space formalism}. We start our analysis in Sec.~\ref{sec:Breit frame distributions} with the 3D Breit frame distributions of the weak-neutral axial-vector four-current densities inside a general spin-$\tfrac{1}{2}$ hadron. We then present in Sec.~\ref{sec:Elastic frame distributions} the generic elastic frame distributions for a moving spin-$\frac{1}{2}$ hadron at arbitrary average momentum. In Sec.~\ref{sec:Light-front distributions}, we present the corresponding light-front distributions, and the explicit demonstration of light-front amplitudes via the proper infinite-momentum frame limit of elastic frame amplitudes. In Sec.~\ref{sec:Breit frame distributions} to Sec.~\ref{sec:Light-front distributions}, we also rederive 3D and 2D transverse mean-square axial and spin radii. Finally, we summarize our findings in Sec.~\ref{sec:Summary}. For completeness, we also provide details for the parametrization of nucleon weak-neutral axial-vector form factors in Appendix~\ref{Appendix-A}, and further discussions on the breakdown of Abel transformation of axial charge distributions in Appendix~\ref{Appendix-B}.

\section{Weak-neutral axial-vector form factors}
\label{sec:Weak-neutral axial-vector form factors}

\begin{figure}[htb!]
	\centering
	{\includegraphics[angle=0,scale=0.5]{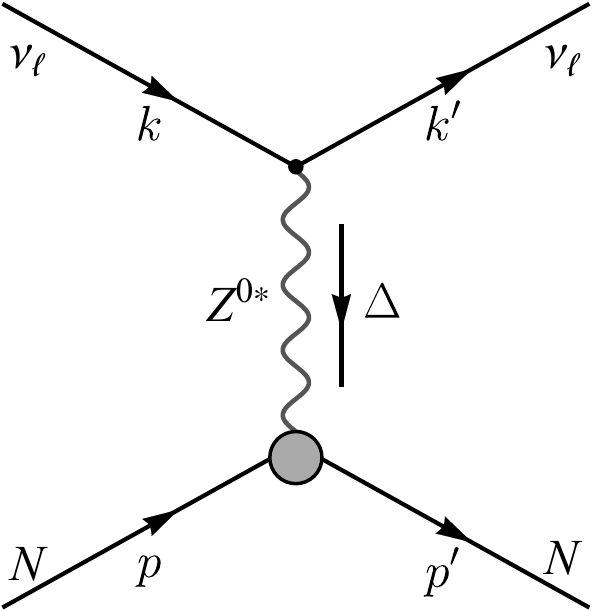}}
	\caption{Illustration of the tree-level Feynman diagram of the $t$-channel weak-neutral current elastic scattering reaction $\nu_\ell(k) + N(p) \to \nu_\ell(k') + N(p')$ in the first Born approximation (i.e. one virtual $Z^0$ boson exchange), with $\ell=e,\mu,\tau$. The four-momentum transfer is $\Delta=k-k'=p'-p$.}
	\label{Fig_FeynmanDiagram}
\end{figure}

Based on solely Lorentz covariance and the hermiticity property of the weak-neutral axial-vector four-current operator $\hat j_{5}^{\mu}(x) \equiv \hat{\bar \psi}(x) \gamma^\mu \gamma^5 \hat\psi(x)$, there are in general three weak-neutral axial-vector FFs of a spin-$\tfrac{1}{2}$ hadron in the weak sector~\cite{Gourdin:1974iq,Ohlsson:1998bk}. These three axial-vector FFs together describe the internal weak-neutral axial-vector content of the system in response to external weak-neutral current interactions; see e.g., Fig.~\ref{Fig_FeynmanDiagram} for the tree-level Feynman diagram of the neutrino-nucleon elastic scattering.

In the case of a spin-$\tfrac{1}{2}$ hadron, matrix elements of the weak-neutral axial-vector four-current operator $\hat j_{5}^{\mu}$ in general can be parametrized as~\cite{Gourdin:1974iq,Ohlsson:1998bk,Bernard:2001rs,Aliev:2007qu}\footnote{We note that matrix elements of the divergence of $\hat j_5^\mu$ have been  investigated recently in Refs.~\cite{Castelli:2024eza,Bhattacharya:2024geo}.}
\begin{equation}
	\begin{aligned}\label{MatrElem-Axial-Z}
		\langle p',s'|\hat j_{5}^{\mu}(0)|p,s\rangle
		&= \bar u(p',s') \Gamma^\mu(P,\Delta) u(p,s),
	\end{aligned}
\end{equation}
with
\begin{equation}\label{vertex-func}
	\Gamma^{\mu}(P,\Delta) \equiv \gamma^\mu \gamma^5 G_A^Z(Q^2) + \frac{\Delta^\mu \gamma^5}{2M} G_P^Z(Q^2) - \frac{\sigma^{\mu\nu} \Delta_\nu \gamma^5 }{2M} G_T^Z(Q^2),
\end{equation}
where $\hat \psi(x)$ denotes the quark field operator (which can either be a flavor-singlet or flavor-multiplet), $P=(p'+p)/2$, $\Delta=p'-p$, and $p^2=p'^2=M^2$ are the on-mass-shell relations for the same hadron of mass $M$ in the initial and final states. Here, $G_A^Z(Q^2)$, $G_P^Z(Q^2)$ and $G_T^Z(Q^2)$ are called axial, induced pseudoscalar and induced pseudotensor FFs, respectively. The hermiticity condition implies that all these three FFs are \emph{real} in the spacelike region as functions of the four-momentum transfer squared $Q^2 \equiv -\Delta^2 \geq 0$. Since the explicit (diagonal) matrix forms of generators $T_a$ associated with the operator $\hat j_{5,a}^{\mu}(x)\equiv \hat{\bar \psi}(x) \gamma^\mu \gamma^5 T_a \hat\psi(x)$, see e.g. Ref.~\cite{Bernard:2001rs}, for a given flavor-space fundamental representation group $SU(n)_f$ ($n \geq 2$ and $n \in \mathbb Z$) of a given (anti)neutrino-hadron elastic scattering reaction are known, it is convenient to work in the $U(1)_f$ representation without loss of generality.

Now, let us take a close look why these weak-neutral axial-vector FFs $G_A^Z(Q^2)$, $G_P^Z(Q^2)$ and $G_T^Z(Q^2)$ are real in Eq~(\ref{vertex-func}). Since $\hat j_5^\mu = ( \hat j_5^\mu )^\dag$, the hermiticity condition implies that
\begin{equation}\label{hermiticity-condition}
	\langle p',s'|\hat j_5^\mu(0)|p,s\rangle = \langle p,s|\hat j_5^\mu(0)|p',s'\rangle^\dag,
\end{equation}
which amounts to $\Gamma^\mu(P,\Delta) = \gamma^0 \Gamma^{\mu,\dag}(P,-\Delta)\gamma^0$, or equivalently $\Gamma^{\mu,\dag}(P,\Delta) = \gamma^0 \Gamma^\mu(P,-\Delta) \gamma^0$. Explicit evaluation of this condition using the full vertex function (\ref{vertex-func}) shows that
\begin{equation}\label{FFs-real}
	G_X^Z(Q^2) = \left[ G_X^Z(Q^2) \right]^*,\qquad X=A, P, T,
\end{equation}
where we have used the following identities for the Dirac $\gamma$ matrices:
\begin{equation}
	\gamma^0 \gamma^\mu \gamma^5 \gamma^0 
	= \gamma^5 (\gamma^\mu)^\dag,\qquad
	\gamma^0 \gamma^5 \gamma^0 
	= -\gamma^5,\qquad
	\gamma^0 \sigma^{\mu\nu} \gamma^5 \gamma^0
	= -\gamma^5 (\sigma^{\mu\nu} )^\dag. 
\end{equation}

According to Weinberg~\cite{Weinberg:1958ut}, one can use the $\mathsf{G}$-parity to classify all possible currents formed by Dirac bilinears in the literature in term of the \emph{first-} and \emph{second-class} currents. The $\mathsf{G}$-parity is defined as the combination of the charge-conjugation $\mathsf{C}$-parity after an rotation of a $180^\circ$ angle around the $y$-axis in isospin space, namely $\mathsf{G} \equiv \mathsf{C}\, \exp\left(  i\pi I_y \right)$, where $I_y= \sigma_y/2 $ is the $y$-component isospin matrix. As the \emph{ne plus ultra}, one can easily demonstrate that the vector and axial-vector four-currents $j^\mu=\bar{\psi}\gamma^\mu \psi$ and $j^\mu_5=\bar{\psi}\gamma^\mu\gamma^5 \psi$ which transform in the following manner
\begin{equation}
	\begin{aligned}\label{G-VA-first}
		\mathsf{G} j^{\mu} \mathsf{G}^{-1} 
		&= +j^{\mu},\qquad
		\mathsf{G} j_5^{\mu} \mathsf{G}^{-1} 
		= -j_5^{\mu}
	\end{aligned}	
\end{equation}
are classified as the \emph{first-class} currents, whereas those transform in the opposite manner
\begin{equation}
	\begin{aligned}\label{G-VA-second}
		\mathsf{G} j^{\mu} \mathsf{G}^{-1} 
		&= -j^{\mu},\qquad
		\mathsf{G} j_5^{\mu} \mathsf{G}^{-1} 
		= +j_5^{\mu}
	\end{aligned}
\end{equation}
are classified as the \emph{second-class} currents.

Using Weinberg's classification~\cite{Weinberg:1958ut}, we can identify $G_A^Z$ and $G_P^Z$ in Eq.~(\ref{vertex-func}) as the FFs associated with the first-class currents, while identify $G_T^Z$ as the FF associated with the second-class current~\cite{Gourdin:1974iq,Ohlsson:1998bk,Fatima:2018tzs}. In the presence of exact isospin symmetry or $\mathsf{G}$-parity invariance, the second-class current contribution vanishes identically~\cite{Ohlsson:1998bk,Bernard:2001rs,Aliev:2007qu}, i.e. $G_T^Z(Q^2)=0$. It is thus common in the literature that the second-class current contribution of a hadron is usually not much discussed or calculated, see e.g. Refs.~\cite{Meissner:1986xg,Meissner:1986js,Bernard:1994wn,Silva:2005fa,Schindler:2006jq,Aliev:2007qu,Sharma:2009hg,Eichmann:2011pv,Liu:2014owa,Dahiya:2014jfa,Ramalho:2015jem,Anikin:2016teg,Mamedov:2016ype,Hashamipour:2019pgy,Mondal:2019jdg,Zhang:2019iyx,Jun:2020lfx,Chen:2020wuq,Chen:2021guo,Ahmady:2021qed,Sauerwein:2021jxb,Xu:2021wwj,Atayev:2022omk,Chen:2022odn,Cheng:2022jxe,Liu:2022ekr,Irani:2023lol,Ramalho:2024tdi}, since the $\mathsf{G}$-parity invariance is well respected by the strong interaction or QCD\footnote{In strong interactions or QCD, e.g. the strong decays of mesons, the $\mathsf{G}$-parity invariance is exact.}. 

However, the $\mathsf{G}$-parity invariance is in general not preserved in electromagnetic and weak interactions, owing to either the quark electric charge or quark mass differences~\cite{Shiomi:1996np}. This is the partial motivation that we wish to take into account the induced pseudotensor FF $G_T^Z(Q^2)$. Another motivation is that following our recent work~\cite{Chen:2024oxx} we wish to figure out whether the 3D nucleon axial charge distribution in the BF is finite or not, and to check whether the nucleon 3D axial (charge) radius in a more general case exists or not, when the full vertex function (\ref{vertex-func}) is taken into account without assuming the $\mathsf{G}$-parity invariance.

\section{Quantum phase-space formalism}
\label{sec:The quantum phase-space formalism}

Although FFs are objects defined in momentum space and extracted from experimental data involving particles with well-defined momenta, their physical interpretation actually resides in position space~\cite{Chen:2023dxp}. Because of Lorentz symmetry, the notion of relativistic spatial distributions in general depends on the target average momentum, hindering therefore in general a probabilistic interpretation in position space. We are therefore naturally led to switch our perspective to a phase-space picture, which is \emph{quasi}probabilistic at the quantum level owing to Heisenberg's uncertainty principle. In this section, we only present the key ingredients of the quantum phase-space formalism, and refer readers to Refs.~\cite{Lorce:2018egm,Lorce:2020onh,Chen:2022smg,Chen:2023dxp} for more details.

In quantum field theory, it has been known for a long time that the expectation value of any an operator $\hat O$ in a physical state $|\Psi\rangle$ can be expressed as~\cite{Wigner:1932eb,Hillery:1983ms,Bialynicki-Birula:1991jwl}
\begin{equation}\label{PS-representation}
	\langle\Psi|\hat O(x)|\Psi\rangle=\sum_{s',s}\int\frac{\ud^3P}{(2\pi)^3}\,\ud^3R\,\rho^{s's}_\Psi(\uvec R,\uvec P)\,\langle\hat O\rangle^{s's}_{\uvec R,\uvec P}(x),
\end{equation}
where $\rho^{s's}_\Psi(\uvec R,\uvec P)$ defines the \emph{Wigner distribution} interpreted as the quantum weight for finding the system at average position $\uvec R=\tfrac{1}{2}(\uvec r'+\uvec r)$ and average momentum $\uvec P=\tfrac{1}{2}(\uvec p'+\uvec p)$. 

Probabilistic densities are then recovered upon integration over either average position or momentum variables~\cite{Lorce:2020onh,Chen:2023dxp}. A compelling feature of the quantum phase-space formalism is that wave-packet details have been cleanly factored out in Eq.~\eqref{PS-representation}. We can then interpret the phase-space amplitude
\begin{equation}\label{internal-distribution}
	\langle\hat O\rangle^{s's}_{\uvec R,\uvec P}(x)=\int\frac{\ud^3\Delta}{(2\pi)^3}\,e^{i\uvec\Delta\cdot\uvec R}\,\frac{\langle P+\tfrac{\Delta}{2},s'|\hat O(x)|P-\tfrac{\Delta}{2},s\rangle}{2\sqrt{p'^0p^0}}
\end{equation}
as the \emph{internal distribution} associated with a state localized in the Wigner sense around average position $\uvec R$ and average momentum $\uvec P$~\cite{Lorce:2018zpf,Lorce:2018egm,Lorce:2021gxs,Lorce:2020onh,Lorce:2022jyi,Lorce:2022cle,Chen:2022smg,Chen:2023dxp,Kim:2021kum,Kim:2022bia,Hong:2023tkv}.

\section{Breit frame distributions}
\label{sec:Breit frame distributions}

The BF is specified by the condition $\uvec P=\uvec 0$. From a phase-space perspective, the BF can be regarded as the average rest frame of the system, where spin structure assumes its simplest form~\cite{Lorce:2020onh,Lorce:2022jyi,Chen:2022odn,Chen:2023dxp,Chen:2024oxx,Lorce:2024ipy,Lorce:2024ewq,Lorce:2025oot}. In this frame, one can obtain fully relativistic 3D spatial distributions of a generic composite system for its static internal structures in the Wigner sense. This is also the reason why so many 3D mean-square radii of a hadron, e.g. (electric) charge $\langle r_\text{ch}^2 \rangle$, mass $\langle r_\text{mass}^2 \rangle$, mechanical $\langle r_\text{mech}^2 \rangle$ and spin $\langle r_\text{spin}^2 \rangle$ radii, are usually defined in this frame~\cite{Gao:2021sml,Polyakov:2018guq,Polyakov:2018zvc,Burkert:2023wzr,Hackett:2023rif,Chen:2023dxp,Chen:2024oxx}.

Since the energy transfer $\Delta^0=\uvec P \cdot \uvec\Delta/P^0$ vanishes automatically in this frame, internal distributions in the BF do not depend on time $x^0$ arising due to the translation invariance of the matrix elements~(\ref{internal-distribution}). Following Refs.~\cite{Lorce:2020onh,Chen:2022smg,Chen:2023dxp,Chen:2024oxx}, relativistic 3D axial-vector four-current distributions in the BF are defined as
\begin{equation}
	\begin{aligned}\label{BF-def}
		J_{5,B}^\mu (\uvec r)
		\equiv \int\frac{\ud^3\Delta}{(2\pi)^3}\, e^{-i\uvec\Delta \cdot \uvec r}\, \frac{\langle p_B', s_B'| \hat j_5^\mu(0)|p_B, s_B \rangle }{2P_B^0}\bigg|_{\uvec P = \uvec 0},
	\end{aligned}
\end{equation}
where $\uvec r = \uvec x -\uvec R$ is the distance relative to the center $\uvec R=\uvec 0$ of the system, $\uvec p_B'=-\uvec p_B = \uvec\Delta/2$, $Q^2=\uvec\Delta^2$, $\tau \equiv Q^2/(4M^2)$, and $P_B^0=p_B^{\prime\, 0}=p_B^0=M\sqrt{1+\tau}$.

\subsection{BF weak-neutral axial-vector four-current distributions}
\label{BF: Weak-neutral axial-vector four-current distributions}

Evaluating the matrix elements (\ref{MatrElem-Axial-Z}) in the BF leads to~\cite{Meissner:1986js,Ohlsson:1998bk,Barquilla-Cano:2002btm,Schindler:2006jq,Chen:2024oxx}
\begin{equation}
	\begin{aligned}\label{BF-amplitudes}
		\mathcal A^0_B
		&= \sqrt{1 + \tau }\,(i\uvec\Delta \cdot \uvec \sigma)_{s_B' s_B} G_T^Z(\uvec\Delta^2),\\
		\uvec{\mathcal A}_B
		&= 2P_B^0 \left[ \uvec\sigma - \frac{ \uvec\Delta (\uvec\Delta \cdot \uvec\sigma ) }{4P_B^0(P_B^0+M) } \right]_{s_B' s_B} G_A^Z(\uvec\Delta^2) - \frac{  \uvec\Delta (\uvec\Delta \cdot \uvec\sigma )_{s_B' s_B} }{ 2M } G_P^Z(\uvec\Delta^2),
	\end{aligned}
\end{equation}
where $\mathcal A_B^\mu \equiv \langle p_B',s_B'|\hat j_{5}^{\mu}(0)|p_B,s_B\rangle$, and explicit canonical polarization indices will be omitted for better legibility hereafter unless necessary.

\begin{figure}[t!]
	\centering
	{\includegraphics[angle=0,scale=0.54]{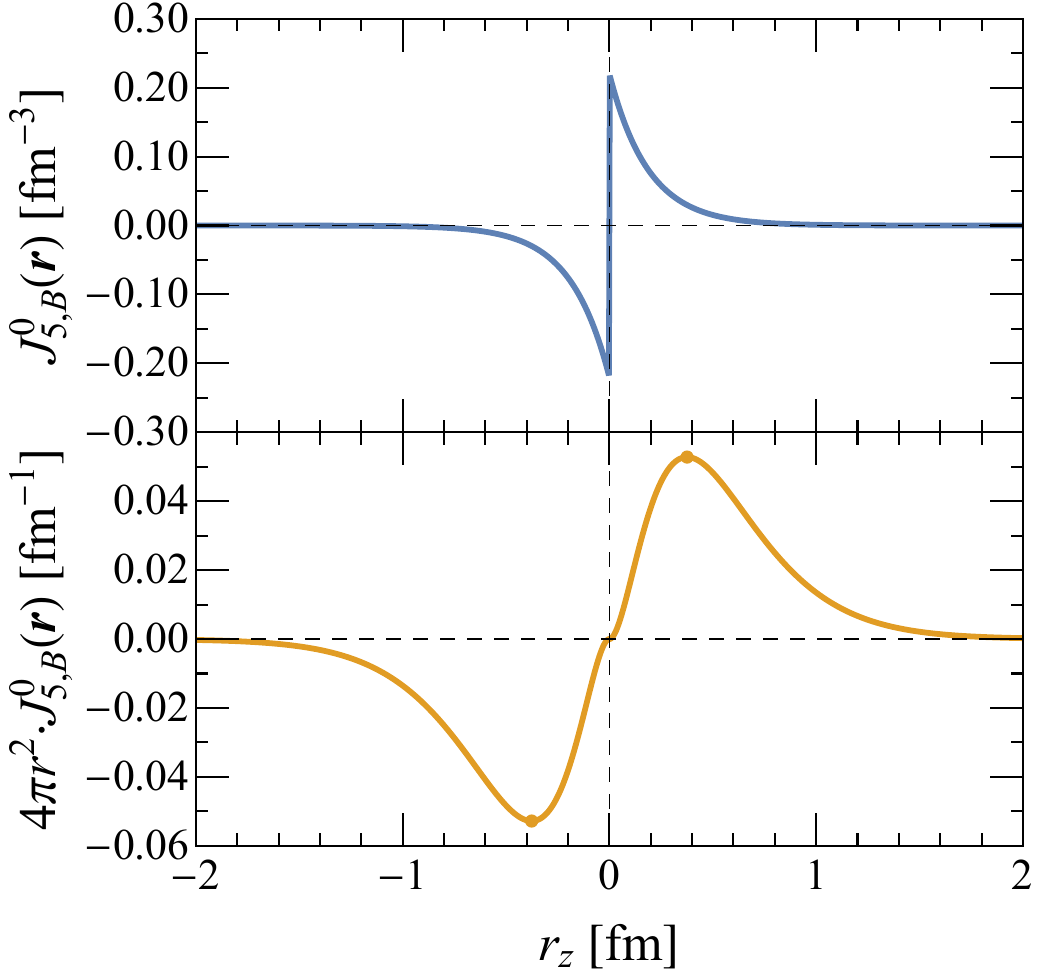}}
	\caption{The 3D weak-neutral axial charge distributions $J_{5,B}^0(\uvec r) $ (upper panel) and $4\pi r^2 \cdot J_{5,B}^0(\uvec r) $ (lower panel) in the BF along the $z$-axis inside a longitudinally polarized (i.e. $\hvec s = \uvec e_z$) proton, using proton's weak-neutral induced pseudotensor FF $G_T^Z(Q^2)$ given in Appendix~\ref{Appendix-A}.}
	\label{Fig_3DBFNC_J0}
\end{figure}

Inserting the BF amplitudes (\ref{BF-amplitudes}) into Eq.~(\ref{BF-def}) leads to the following relativistic 3D BF weak-neutral axial charge and axial-vector current distributions~\cite{Chen:2024oxx}
\begin{equation}
	\begin{aligned}\label{3DBF-distributions}
		J_{5,B}^0(\uvec r)
		&= \int\frac{\ud^3\Delta}{(2\pi)^3}\, e^{-i\uvec\Delta \cdot \uvec r}\, \frac{ i\uvec\Delta \cdot \uvec \sigma}{2M} G_T^Z(\uvec\Delta^2) = -\uvec\nabla_{\uvec r} \cdot \frac{ \uvec\sigma }{2M}\int\frac{\ud^3\Delta}{(2\pi)^3}\, e^{-i\uvec\Delta \cdot \uvec r}\, G_T^Z(\uvec\Delta^2),\\
		\uvec J_{5,B}(\uvec r)
		&= \int\frac{\ud^3\Delta}{(2\pi)^3}\, e^{-i\uvec\Delta \cdot \uvec r}\, \bigg\{ \left[ \uvec\sigma - \frac{ \uvec\Delta (\uvec\Delta \cdot \uvec\sigma ) }{4P_B^0(P_B^0+M) } \right] G_A^Z(\uvec\Delta^2) - \frac{\uvec\Delta (\uvec\Delta \cdot \uvec\sigma ) }{ 4M P_B^0 } G_P^Z(\uvec\Delta^2) \bigg\}.
	\end{aligned}
\end{equation}
We find that the 3D axial charge distribution $J_{5,B}^0(\uvec r)$ is in fact related to the weak-neutral induced pseudotensor FF $G_T^Z(Q^2)$ rather than the axial FF $G_A^Z(Q^2)$. We stress that above results (\ref{3DBF-distributions}) are also well consistent with our previous work~\cite{Chen:2024oxx}, where the $\mathsf{G}$-parity invariance of QCD is further applied to the matrix elements (\ref{MatrElem-Axial-Z}), eliminating therefore the second-class current contribution associated with $G_T^Z(Q^2)$~\cite{Ohlsson:1998bk,Bernard:2001rs,Aliev:2007qu}. Besides, we notice that $J_{5,B}^0(\uvec r)$ is parity-odd owing to the associated parity-odd factor $( \uvec r \cdot \uvec\sigma)$ coming from the Fourier transform of $(\uvec \Delta \cdot \uvec\sigma)$. In Fig.~\ref{Fig_3DBFNC_J0}, we illustrate the parity-odd 3D axial charge distributions $J_{5,B}^0(\uvec r) $ and $4\pi r^2 \cdot J_{5,B}^0(\uvec r) $ with $r=|\uvec r|$ along the $z$-axis inside a longitudinally polarized (i.e. the unit polarization vector\footnote{In general, the unit polarization vector is given by $\hvec s = \chi_{h'}^\dag \uvec\sigma \chi_h = \uvec\sigma_{h'h}$.} $\hvec s = \uvec e_z$) proton, using proton's weak-neutral induced pseudotensor FF given in Appendix~\ref{Appendix-A}.

\begin{figure}[t!]
	\centering
	{\includegraphics[angle=0,scale=0.52]{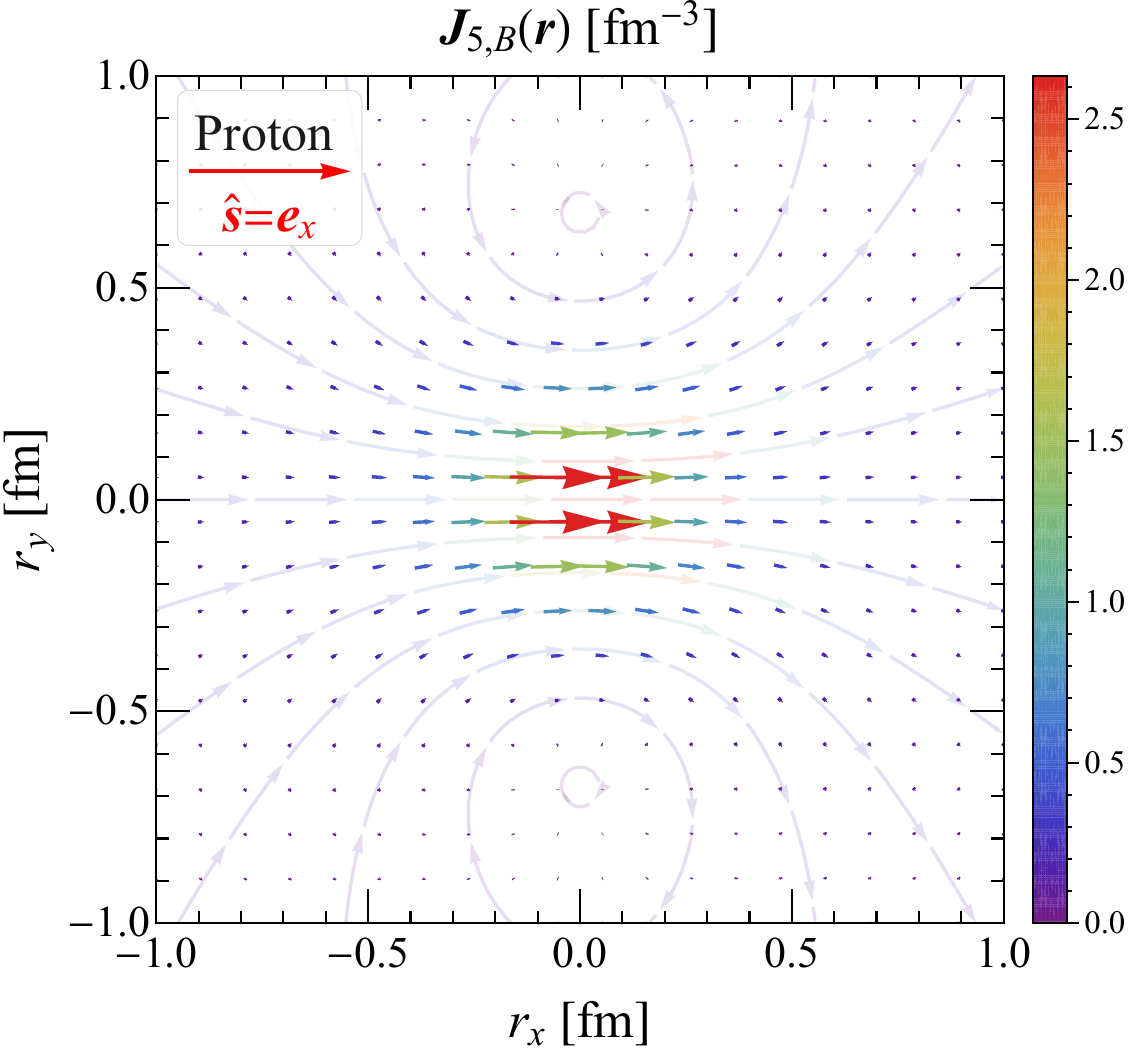}}
	\caption{The 3D weak-neutral axial-vector current distribution $\uvec J_{5,B}(\uvec r)$ in the BF in the transverse plane inside a transversely polarized (i.e. $\hvec s = \uvec e_x$) proton, using proton's weak-neutral axial-vector FFs $G_A^Z(Q^2)$ and $G_P^Z(Q^2)$ given in Appendix~\ref{Appendix-A}.}
	\label{Fig_3DBFNC_Jv}
\end{figure}

Moreover, we do notice that the 3D axial-vector current distribution $\uvec J_{5,B}$ (\ref{3DBF-distributions}) is in fact independent of the induced pseudotensor FF $G_T^Z(Q^2)$. In other words, $\uvec J_{5,B}$ is free from the second-class current contribution, and therefore assumes the same expression as in Ref.~\cite{Chen:2024oxx}. Using the QCD equation of motion~\cite{Leader:2013jra}, one can show that $\uvec S_B(\uvec r) = \uvec J_{5,B}(\uvec r)/2$ is the physically meaningful 3D (intrinsic) spin distribution in the BF~\cite{Lorce:2017wkb,Lorce:2018egm,Lorce:2022cle,Chen:2024oxx}, characterizing how the spin is distributed in the weak-neutral sector. In Fig.~\ref{Fig_3DBFNC_Jv}, we illustrate the 3D weak-neutral axial-vector current distribution $\uvec J_{5,B}(\uvec r)$ in the transverse plane inside a transversely polarized proton, using proton's weak-neutral axial-vector FFs given in Appendix~\ref{Appendix-A}. We observe that the distribution is perfectly mirror symmetric (antisymmetric) with respect to the $x$-axis ($y$-axis) in the transverse plane, exhibiting the so-called toroidal mode~\cite{Repko:2012rj,vonNeumann-Cosel:2023aep} in the weak sector. This can be more easily understood from the multipole decomposition~\cite{Chen:2023dxp} of $\uvec J_{5,B}(\uvec r)$. In doing so, we find that $\uvec J_{5,B}(\uvec r)$ consists of mirror symmetric (antisymmetric) monopole and quadrupole contributions solely with respect to the $x$-axis ($y$-axis).

Recalling in the BF that $\uvec p_B'=-\uvec p_B=\uvec\Delta/2$ and $p_B^{\prime\, 0}=p_B^0 =P_B^0$, we do recognize in Eq.~(\ref{BF-amplitudes}) the characteristic spin structure $\left(\uvec\sigma - \frac{\uvec p_B(\uvec p_B\cdot \uvec\sigma)}{p_B^0(p_B^0+M)} \right)$ defined relative to the center of mass $\uvec R_M $, which is the only relativistic center transforming as the spatial part of a Lorentz four-vector, and corresponds therefore to a physically more natural and transparent relativistic center of the system~\cite{Lorce:2018zpf,Lorce:2021gxs,Chen:2023dxp}. There are other choices of the relativistic center~\cite{Lorce:2018zpf,Lorce:2021gxs,Chen:2023dxp}, e.g. the center of energy $\uvec R_E=\uvec R_M + \frac{\uvec P \times \hvec s}{2M(E_P+M)} $ and the center of (canonical) spin $\uvec R_c=\uvec R_M + \frac{\uvec P \times \hvec s}{2M E_P}$ with $E_P \equiv \sqrt{\uvec P^2 + M^2} $. Different from $\uvec R_M$, $\uvec R_E$ and $\uvec R_c$ however will cause respectively sideways shifts $\frac{\uvec P \times \hvec s}{2M(E_P+M)}$ and $\frac{\uvec P \times \hvec s}{2M E_P}$ of pure relativistic origin when $(\uvec P \times \hvec s) \neq \uvec 0$, e.g, when the spinning system is longitudinally moving while it is transversely polarized. This in turn justifies that the parametrization (\ref{MatrElem-Axial-Z}) is indeed physically clear and transparent, since the spin is defined relative to the center of mass. For more details of the relativistic centers and sideways shifts, see Refs.~\cite{Lorce:2018zpf,Lorce:2021gxs} and also the similar discussions for the parametrization of polarization-magnetization tensor in Ref.~\cite{Chen:2023dxp}.

\subsection{BF mean-square radii}
\label{BF: Mean-square radii}

Although the induced pseudotensor FF $G_T^Z(Q^2)$ is explicitly taken into account (\ref{vertex-func}), we note that the total axial charge of a general spin-$\frac{1}{2}$ composite system in the BF vanishes identically~\cite{Chen:2024oxx}:
\begin{equation}
	\begin{aligned}
		\int \ud^3 r\, J_{5,B}^0(\uvec r) = \lim_{\uvec\Delta \to \uvec 0}\, \frac{1}{2M}\left[ (i\uvec\Delta \cdot \uvec\sigma) G_T^Z(\uvec\Delta^2) \right] = 0,
	\end{aligned}
\end{equation}
because of the parity-odd nature of the 3D axial charge distribution $J_{5,B}^0(\uvec r)$ itself. This means that the definition of the standard mean-square axial (charge) radius 
\begin{equation}
	\begin{aligned}\label{3DBF-MS-axial-radius}
		\langle r_A^2 \rangle
		&\equiv \frac{\int \ud^3r\, r^2\,J_{5,B}^0(\uvec r) }{ \int \ud^3r\, J_{5,B}^0(\uvec r)  }
	\end{aligned}
\end{equation}
for a general spin-$\frac{1}{2}$ hadron is in fact \emph{not} well-defined. This is different from case for the definition of the mean-square charge radius of the neutron where one can replace $G_E^n(0)=0$ with $G_E^p(0)=1$~\cite{Gao:2021sml,Chen:2023dxp} so as to make the definition well-defined\footnote{Note also the fact that the BF charge distribution of the neutron is spherically symmetric~\cite{Lorce:2020onh,Chen:2023dxp}.}, since the induced pseudotensor charge $G_T^Z(0)$ is in general not zero~\cite{Day:2012gb}, see e.g., Eq.~(\ref{NC-Z-FF-GT}), while the distribution itself is parity-odd.

In the case $G_T^Z(Q^2)=0$ by using the $\mathsf G$-parity invariance of QCD, $J_{5,B}^0(\uvec r)$ vanishes identically and thus the axial (charge) radius does not exist~\cite{Chen:2024oxx}, contrary to what is usually stated in the literature~\cite{Meissner:1986xg,Meissner:1986js,Bernard:1992ys,A1:1999kwj,Meyer:2016oeg,Hill:2017wgb,MINERvA:2023avz,Petti:2023abz,Kaiser:2024vbc} via
\begin{equation}
	\begin{aligned}\label{naive-3DMS-axial-radius}
		R_A^2 \equiv -\frac{6}{G_A^Z(0) }\frac{\ud G_A^Z(Q^2) }{\ud Q^2} \bigg|_{Q^2=0} = \frac{1}{G_A^Z(0)}\left[ -\uvec\nabla_{\uvec\Delta}^2 G_A^Z(\uvec\Delta^2) \right] \Big|_{\uvec\Delta = \uvec 0}.
	\end{aligned}
\end{equation}
In conclusion, in either $G_T^Z(Q^2) = 0$ or $G_T^Z(Q^2) \neq 0$ cases, the genuine 3D mean-square axial (charge) radius $\langle r_A^2 \rangle $ is not defined via Eq.~(\ref{naive-3DMS-axial-radius}), since the 3D axial charge distribution itself is related to $G_T^Z(Q^2)$ rather than $G_A^Z(Q^2)$.

On the other hand, the 3D mean-square spin radius $\langle r_\text{spin}^2 \rangle$ is a well-defined and physically meaningful quantity~\cite{Chen:2024oxx}. Using proton's weak-neutral axial-vector FFs given in Appendix~\ref{Appendix-A} and the same formula for the mean-square spin radius given in Refs.~\cite{Chen:2024oxx,Chen:2024ksq}, we find for the proton that $\langle r_\text{spin}^2 \rangle \approx (2.1054~\text{fm})^2$ and $R_A^2 \approx (0.6510~\text{fm})^2$. We reconfirm that the relativistic contribution $\frac{1}{4M^2}\left(1 + \frac{2G_P^Z(0) }{G_A^Z(0)} \right) $ indeed plays a dominant role~\cite{Chen:2024oxx}.

Before we move forward, there are two key points deserving to be emphasized. The first point is that the relation between a genuine 3D mean-square radius and the slope of the corresponding FF is in general not so obvious and simple, see the examples given in Refs.~\cite{Polyakov:2018zvc,Chen:2024oxx}. One needs to carefully define first the corresponding distribution, and then derive the genuine mean-square radius based on that distribution. The second point is that the concept of identifying a mean-square radius simply via the slope of the FF with same naming is in general misleading and incorrect. A typical example is the 3D mean-square axial (charge) radius for which the 3D axial charge distribution $J_{5,B}^0(\uvec r)$ is not even related to the axial FF $G_A^Z(Q^2)$, see Eq.~(\ref{3DBF-distributions}) and Ref.~\cite{Chen:2024oxx}. Furthermore, our result of the axial charge distribution (\ref{3DBF-distributions}) also explicitly reveals the breakdown of Abel transformation, see the further discussions given in Appendix~\ref{Appendix-B}.

\section{Elastic frame distributions}
\label{sec:Elastic frame distributions}

BF distributions provide us the best proxy for picturing a
system in 3D sitting in average at rest around the origin in the Wigner sense, where the spin in the parametrization (\ref{MatrElem-Axial-Z}) is defined relative to the relativistic center of mass. If one is however interested in the internal structure of a moving system, one can employ the so-called elastic frame (EF) distributions introduced in Ref.~\cite{Lorce:2017wkb}.

Following Refs.~\cite{Lorce:2020onh,Chen:2023dxp,Chen:2024oxx}, the relativistic axial-vector four-current distributions for a moving target in the generic EF are defined as
\begin{equation}\label{EF-def}
	\begin{aligned}
		J_{5,\text{EF}}^\mu(\uvec b_\perp;P_z)
		&\equiv \int \frac{\ud^2 \Delta_\perp}{(2\pi)^2}\, e^{-i\uvec\Delta_\perp \cdot \uvec b_\perp }\, \frac{\langle p',s'| \hat j_5^\mu(0) |p,s \rangle }{2P^0} \bigg|_{\Delta_z=|\uvec P_\perp|=0},
	\end{aligned}
\end{equation}
where the $z$-axis has been chosen along $\uvec P=(\uvec 0_\perp, P_z)$ without loss of generality. Since the energy transfer $\Delta^0 = \uvec P \cdot \uvec \Delta/P^0 $ vanishes automatically, EF distributions (\ref{EF-def}) are indeed independent of time.

\subsection{EF weak-neutral axial-vector four-current distributions}
\label{EF: Weak-neutral axial-vector four-current distributions}

Evaluating directly the matrix elements (\ref{MatrElem-Axial-Z}) in the generic EF leads to~\cite{Chen:2024oxx}
\begin{equation}
	\begin{aligned}\label{EF-amplitudes}
		\mathcal A_\text{EF}^0
		&= 2P^0 \left[ \frac{(i\uvec\Delta_\perp \cdot \uvec\sigma_\perp)}{2M} G_T^Z(\uvec\Delta_\perp^2) + \frac{P_z}{P^0} \sigma_z G_A^Z(\uvec\Delta_\perp^2) \right],\\
		\mathcal A_\text{EF}^z
		&= 2P^0 \left[ \frac{P_z}{P^0} \frac{ (i\uvec\Delta_\perp \cdot \uvec\sigma_\perp) }{2 M}  G_T^Z(\uvec\Delta_\perp^2) +  \sigma_z G_A^Z(\uvec\Delta_\perp^2) \right],\\
		\uvec{\mathcal A}_{\text{EF}}^\perp
		&= 2\sqrt{P^2} \left[ \frac{ P^0+M(1+\tau) }{ (P^0+M)\sqrt{1+\tau} } \uvec\sigma_\perp + \frac{(\uvec e_z \times i\uvec\Delta_\perp)_\perp }{2M} \frac{P_z }{ (P^0+M)\sqrt{1+\tau} } \right] G_A^Z(\uvec\Delta_\perp^2)\\
		&\qquad- \frac{\uvec\Delta_\perp( \uvec\Delta_\perp \cdot \uvec\sigma_\perp) }{2}\left[ \frac{ G_A^Z(\uvec\Delta_\perp^2) }{ P^0+M } + \frac{ G_P^Z(\uvec\Delta_\perp^2) }{ M }  \right] ,
	\end{aligned}
\end{equation}
where $\mathcal A_\text{EF}^\mu \equiv \langle p',s'|\hat j_{5}^{\mu}(0)|p,s\rangle$, $p'=(P^0, \uvec\Delta_\perp/2, P_z)$, $p=(P^0, -\uvec\Delta_\perp/2, P_z)$, $Q^2=\uvec\Delta_\perp^2$, and $P^0 = \sqrt{M^2(1+\tau)+P_z^2}$.

\begin{figure}[b!]
	\centering
	{\includegraphics[angle=0,scale=0.53]{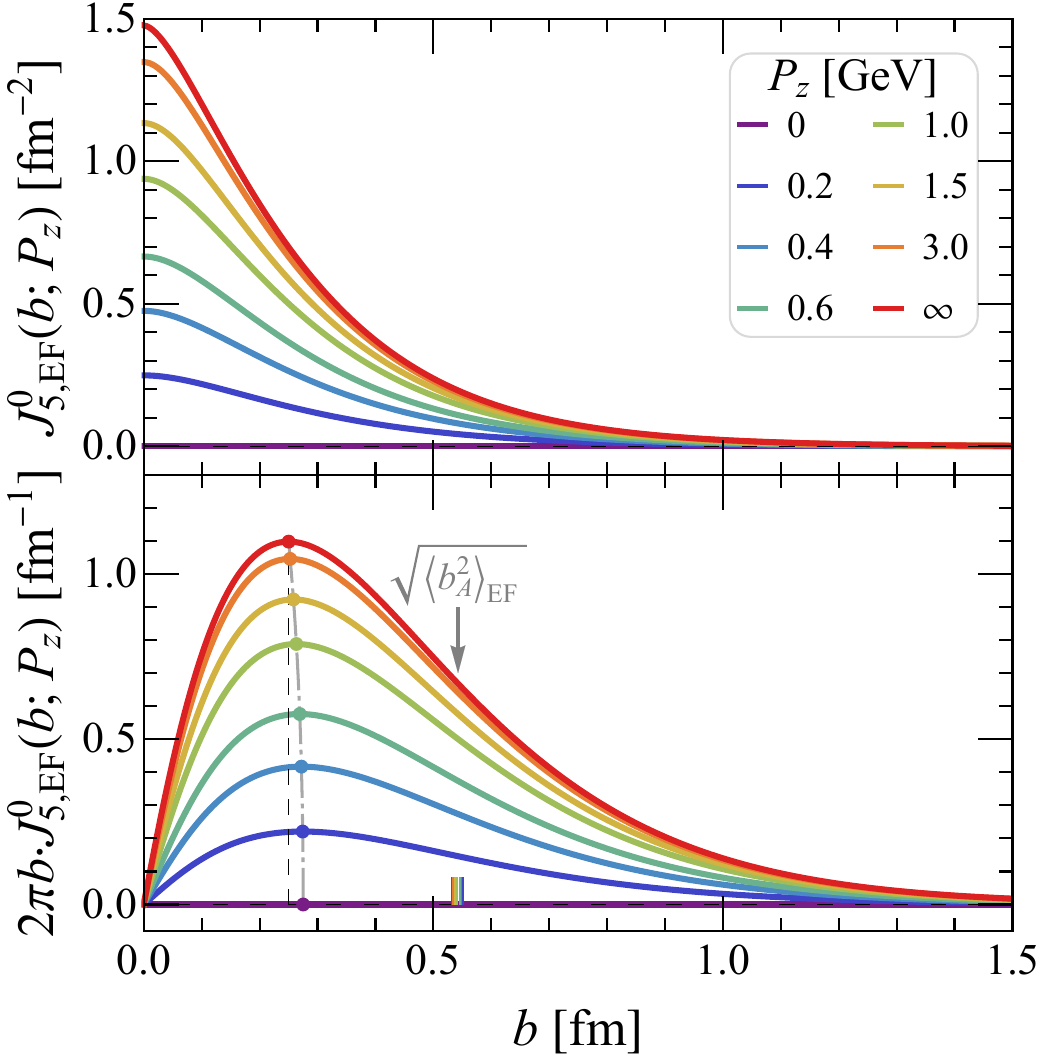}}
	\caption{Frame-dependence of EF weak-neutral axial charge distributions $J_{5,\text{EF}}^0(b;P_z) $ (upper panel) and $2\pi b \cdot J_{5,\text{EF}}^0(b;P_z) $ (lower panel) as a function $b=|\uvec b_\perp|$ inside a longitudinally polarized (i.e. $\hvec s = \uvec e_z$) proton, using proton's weak-neutral axial FF $G_A^Z(Q^2)$ given in Appendix~\ref{Appendix-A}. Maxima of $2\pi b \cdot J_{5,\text{EF}}^0$ are indicated by the gray dot-dashed curve, and the corresponding root-mean-square transverse axial (charge) radii $\sqrt{\langle b_A^2 \rangle_\text{EF}}$ of $J_{5,\text{EF}}^0$ from Eq.~(\ref{EF-MS-transverse-radii}) are also marked in the lower panel.}
	\label{Fig_2DEFNC_J0Sz}
\end{figure}

Poincar\'e symmetry can also be employed to determine how the matrix elements of the axial-vector four-current operator in different Lorentz frames are related to each other. One can write in general~\cite{Jacob:1959at,Durand:1962zza}
\begin{equation}
	\begin{aligned}\label{covariant-Lorentz-Transf}
		\langle p',s'|\hat j_5^\mu(0)|p,s\rangle 
		&= \sum_{s_{B}',s_{B} } D^{\dag(j)}_{s' s_{B}'}(p_{B}',\Lambda) D^{(j)}_{s_{B}s}(p_{B},\Lambda)\, \Lambda^{\mu}_{\phantom{\mu}\nu} \, \langle p_{B}',s_{B}'|\hat j_5^{\nu}(0)|p_{B},s_{B}\rangle,
	\end{aligned}
\end{equation}
where $\langle p_{B}',s_{B}'|\hat j_5^{\nu}(0)|p_{B},s_{B}\rangle$ represent the BF matrix elements, $\Lambda$ is the rotationless Lorentz boost matrix from the BF to a generic Lorentz frame, and $D^{(j)}$ is the Wigner rotation matrix for spin-$j$ systems. Alternatively, one can \emph{analytically} reproduce above EF amplitudes (\ref{EF-amplitudes}) by applying the covariant Lorentz transformation (\ref{covariant-Lorentz-Transf}) on the BF amplitudes (\ref{BF-amplitudes}) at $\Delta_z=0$, with the help of Wigner rotation matrix $D^{(1/2)}$ given in Ref.~\cite{Chen:2023dxp}. In doing so, we do explicitly reproduce our previous results~\cite{Lorce:2022jyi,Chen:2022smg,Chen:2023dxp,Chen:2024oxx} for the Wigner angular conditions:
\begin{equation}
	\begin{aligned}\label{Wigner-angles}
		\cos\theta
		&= \frac{ P^0 + M(1+\tau) }{ (P^0+M)\sqrt{1+\tau} },\qquad
		\sin\theta
		= -\frac{\sqrt{\tau} P_z }{ (P^0+M)\sqrt{1+\tau} }.
	\end{aligned}
\end{equation}
As a result, EF amplitudes (\ref{EF-amplitudes}) can be equivalently rewritten as~\cite{Chen:2024oxx}
\begin{equation}
	\begin{aligned}\label{EF-amplitudes-2}
		\mathcal A_\text{EF}^0
		&= 2 M\gamma \sqrt{1+\tau} \left[  \frac{ (i \uvec\Delta_\perp \cdot \uvec\sigma_\perp) }{2M}  G_T^Z(\uvec\Delta_\perp^2) + \beta \sigma_z G_A^Z(\uvec\Delta_\perp^2) \right],\\
		\mathcal A_\text{EF}^z
		&= 2 M\gamma \sqrt{1+\tau} \left[ \beta \frac{ (i \uvec\Delta_\perp \cdot \uvec\sigma_\perp) }{2M}  G_T^Z(\uvec\Delta_\perp^2) + \sigma_z G_A^Z(\uvec\Delta_\perp^2) \right],\\
		\uvec{\mathcal A}_{\text{EF}}^\perp
		&= 2M\sqrt{1+\tau} \bigg\{ \left[ \cos\theta\, \uvec\sigma_\perp - \frac{(\uvec e_z \times i\uvec\Delta_\perp)_\perp }{ 2M\sqrt{\tau} } \sin\theta  \right] G_A^Z(\uvec\Delta_\perp^2) \\
		&\qquad -\frac{\uvec\Delta_\perp( \uvec\Delta_\perp \cdot \uvec\sigma_\perp) }{ 4M\sqrt{1+\tau} }\left[ \frac{ G_A^Z(\uvec\Delta_\perp^2) }{ P^0+M } + \frac{ G_P^Z(\uvec\Delta_\perp^2) }{ M }  \right] \bigg\},
	\end{aligned}
\end{equation}
with $\gamma \equiv P^0/\sqrt{P^2} $ and $\beta \equiv P_z/P^0$. Inserting the EF amplitudes (\ref{EF-amplitudes}) or (\ref{EF-amplitudes-2}) into Eq.~(\ref{EF-def}) leads to the following relativistic EF weak-neutral axial-vector four-current distributions:
\begin{equation}
	\begin{aligned}\label{2DEF-distributions}
		J_{5,\text{EF}}^0(\uvec b_\perp;P_z)
		&= \int \frac{\ud^2 \Delta_\perp}{(2\pi)^2}\, e^{-i\uvec\Delta_\perp \cdot \uvec b_\perp }\, \left[ \frac{ (i\uvec\Delta_\perp \cdot \uvec\sigma_\perp)_{s's} }{2M} G_T^Z(\uvec\Delta_\perp^2) + \left( \frac{P_z}{P^0}\right) (\sigma_z)_{s's}  G_A^Z(\uvec\Delta_\perp^2) \right] ,\\
		J_{5,\text{EF}}^z(\uvec b_\perp;P_z)
		&= \int \frac{\ud^2 \Delta_\perp}{(2\pi)^2}\, e^{-i\uvec\Delta_\perp \cdot \uvec b_\perp }\, \left[ \left( \frac{P_z}{P^0} \right) \frac{ (i\uvec\Delta_\perp \cdot \uvec\sigma_\perp)_{s's} }{2M} G_T^Z(\uvec\Delta_\perp^2) + (\sigma_z)_{s's} G_A^Z(\uvec\Delta_\perp^2) \right] ,\\
		\uvec J_{5,\text{EF}}^\perp(\uvec b_\perp;P_z)
		&= \int\frac{\ud^2\Delta_\perp }{(2\pi)^2 }\, e^{-i\uvec\Delta_\perp \cdot \uvec b_\perp }\, \bigg\{ - \frac{\uvec\Delta_\perp( \uvec\Delta_\perp \cdot \uvec\sigma_\perp)_{s's} }{ 4P^0 }\left[ \frac{ G_A^Z(\uvec\Delta_\perp^2) }{ P^0+M } + \frac{ G_P^Z(\uvec\Delta_\perp^2) }{ M }  \right]\\
		&\qquad + \frac{\sqrt{P^2}}{P^0} \left[ \cos\theta\, (\uvec\sigma_\perp)_{s's} - \frac{(\uvec e_z \times i\uvec\Delta_\perp)_\perp }{ 2M\sqrt{\tau} } \sin\theta\, \delta_{s' s} \right] G_A^Z(\uvec\Delta_\perp^2) \bigg\}.
	\end{aligned}
\end{equation}
We remind that $\uvec S_\text{EF}(\uvec b_\perp;P_z) = \uvec J_{5,\text{EF}}(\uvec b_\perp;P_z)/2$ is the EF spin distribution~\cite{Lorce:2017wkb,Lorce:2018egm,Lorce:2022cle,Chen:2024oxx}. In Fig.~\ref{Fig_2DEFNC_J0Sz}, we illustrate the EF axial charge distributions $J_{5,\text{EF}}^0(b;P_z)$ and $2\pi b \cdot J_{5,\text{EF}}^0(b;P_z)$ as a function of $b=|\uvec b_\perp|$ inside a longitudinally polarized (i.e. $\hvec s = \uvec e_z$) proton at different $P_z$, using proton's weak-neutral axial FF given in Appendix~\ref{Appendix-A}. We see that $J_{5,\text{EF}}^0$ in Fig.~\ref{Fig_2DEFNC_J0Sz} strongly depends on $P_z$ and vanishes identically when $P_z=0$, reflecting the fact that it is an induced effect for a longitudinally polarized spin-$\frac{1}{2}$ target. This observation is also consistent with the fact that $J_{5,B}^0$ is entirely contributed by $G_T^Z(Q^2)$ in the BF (\ref{3DBF-distributions}).

\begin{figure}[t!]
	\centering
	{\includegraphics[angle=0,scale=0.56]{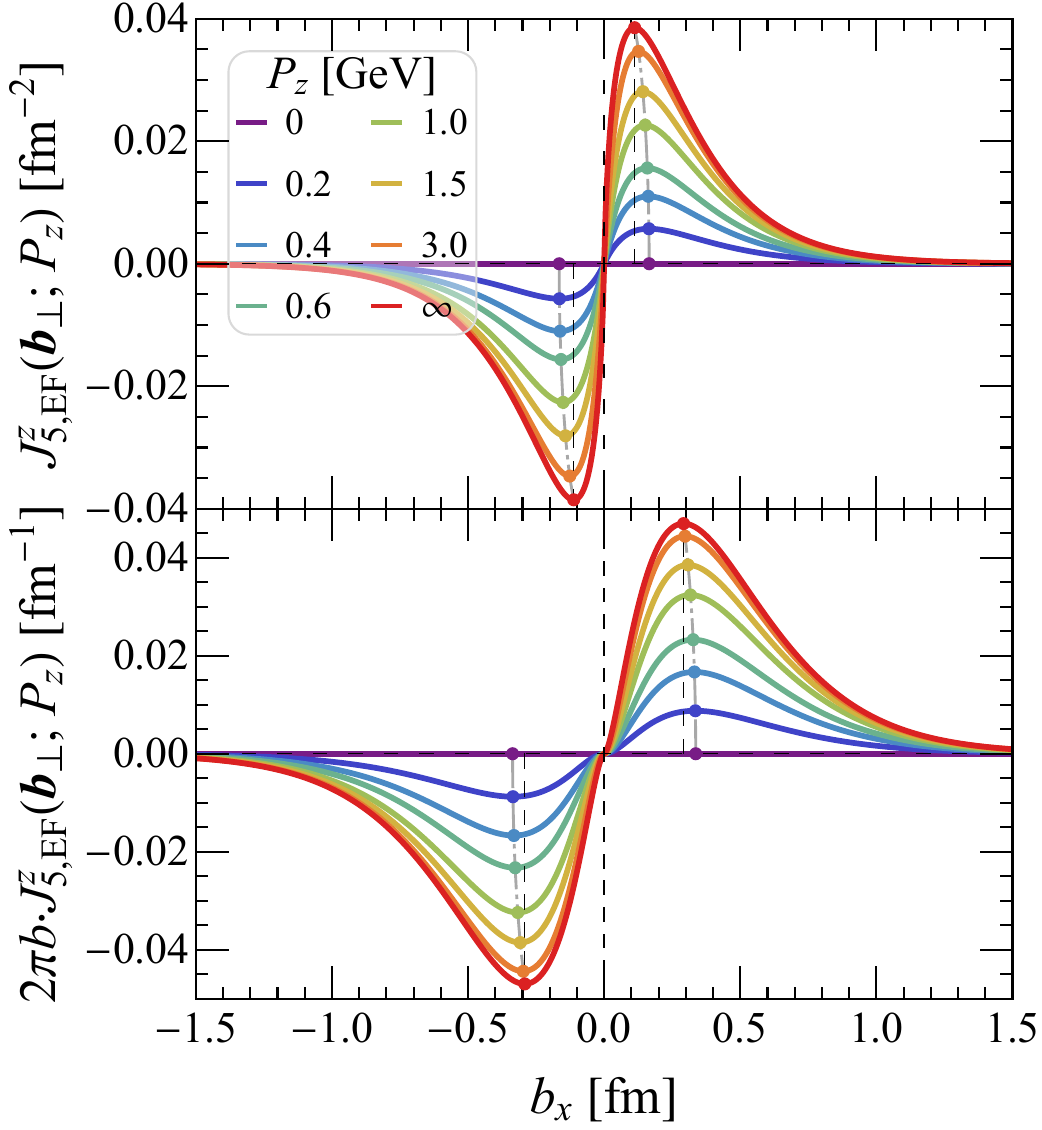}}
	\caption{Frame-dependence of EF weak-neutral longitudinal axial-vector current distributions $J_{5,\text{EF}}^z(\uvec b_\perp;P_z) $ (upper panel) and $2\pi b \cdot J_{5,\text{EF}}^z(\uvec b_\perp;P_z)$ (lower panel) along the $x$-axis inside a transversely polarized (i.e. $\hvec s = \uvec e_x$) proton, using proton's weak-neutral induced pseudotensor FF $G_T^Z(Q^2)$ given in Appendix~\ref{Appendix-A}. The maxima (minima) are indicated by the gray dot-dashed curves.}
	\label{Fig_2DEFNC_JzSx}
\end{figure}

According to Eq.~(\ref{2DEF-distributions}), it is clear that $J_{5,\text{EF}}^0(\uvec b_\perp;\infty) = J_{5,\text{EF}}^z(\uvec b_\perp;\infty)$, which is quite reminiscent of the result in the electromagnetic case~\cite{Chen:2022smg,Chen:2023dxp}. In Fig.~\ref{Fig_2DEFNC_JzSx}, we illustrate the longitudinal axial-vector current distributions $J_{5,\text{EF}}^z(\uvec b_\perp;P_z) $ and $2\pi b \cdot J_{5,\text{EF}}^z(\uvec b_\perp;P_z)$ along the $x$-axis at different $P_z$ inside a transversely polarized proton, using proton's weak-neutral induced pseudotensor FF given in Appendix~\ref{Appendix-A}. We see that the distributions in Fig.~\ref{Fig_2DEFNC_JzSx} are indeed parity-odd similar as Fig.~\ref{Fig_3DBFNC_J0}, and have strong $P_z$-dependence. We note that if the proton is longitudinally polarized, $J_{5,\text{EF}}^z$ will be axially symmetric and frame-independent, and it coincides exactly with the $P_z=\infty$ case of $J_{5,\text{EF}}^0(b;P_z)$ in Fig.~\ref{Fig_2DEFNC_J0Sz}.

\begin{figure}[t!]
	\centering
	{\includegraphics[angle=0,scale=0.52]{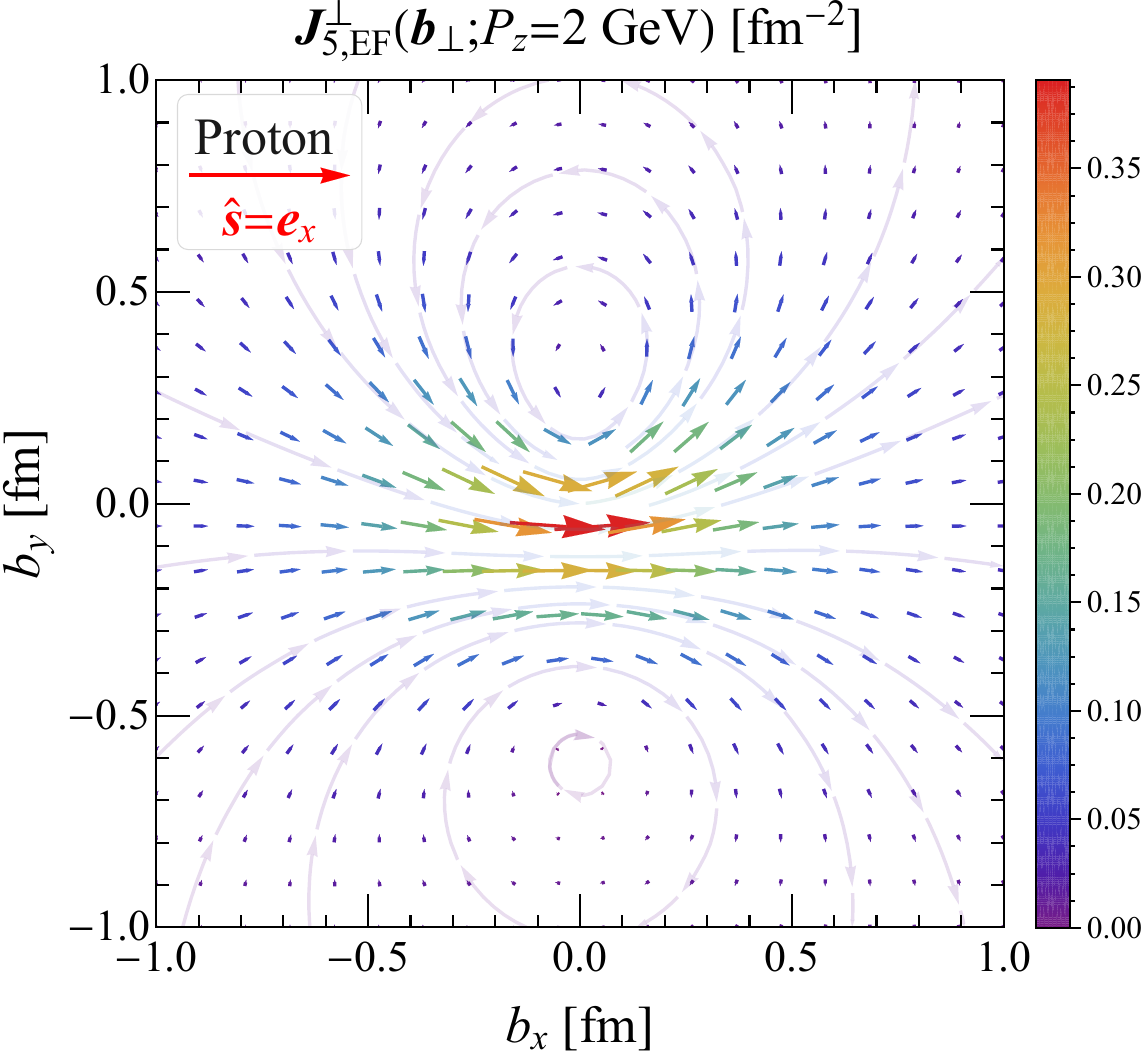}}
	\caption{The transverse EF weak-neutral axial-vector current distribution $\uvec J_{5,\text{EF}}^\perp(\uvec b_\perp;P_z)$ in the transverse plane inside a transversely polarized (i.e. $\hvec s = \uvec e_x$) proton at $P_z=2~\text{GeV}$, using proton's weak-neutral axial-vector FFs $G_A^Z(Q^2)$ and $G_P^Z(Q^2)$ given in Appendix~\ref{Appendix-A}.}
	\label{Fig_2DEFNC_JvT}
\end{figure}

We do observe that the transverse EF axial-vector current distribution $\uvec J_{5,\text{EF}}^\perp$ is independent of $G_T^Z(Q^2)$, since the transverse part of the BF axial-vector four-current amplitudes (\ref{BF-amplitudes}) does not get mixed under longitudinal Lorentz boosts (\ref{covariant-Lorentz-Transf}). In Fig.~\ref{Fig_2DEFNC_JvT}, we illustrate the transverse EF weak-neutral axial-vector current distribution $\uvec J_{5,\text{EF}}^\perp(\uvec b_\perp;P_z)$ in the transverse plane inside a transversely polarized proton at $P_z=2~\text{GeV}$, using proton's weak-neutral axial-vector FFs given in Appendix~\ref{Appendix-A}. We observe that the EF distribution $\uvec J_{5,\text{EF}}^\perp$ at $P_z = 2~\text{GeV}$, compared with the BF one at $P_z = 0$ in Fig.~\ref{Fig_3DBFNC_Jv}, is no longer mirror symmetric with respect to the $x$-axis but is still mirror antisymmetric with respect to the $y$-axis in the transverse plane, and it does exhibit the toroidal mode~\cite{Repko:2012rj,vonNeumann-Cosel:2023aep} similar as Fig.~\ref{Fig_3DBFNC_Jv}. The key reason is that on top of the monopole and quadrupole contributions, $\uvec J_{5,\text{EF}}^\perp$ at finite $P_z$ also contains a dipole contribution, which explicitly breaks the up-down mirror symmetry (with respect to the $x$-axis) but still preserves the left-right mirror antisymmetry (with respect to the $y$-axis) in the transverse plane for a transversely polarized spin-$\frac{1}{2}$ target.

Furthermore, owing to the Lorentz mixing of temporal and longitudinal components of the axial-vector four-current amplitudes under a longitudinal Lorentz boost, we also notice that as long as $P_z$ is nonvanishing the generic EF distributions $J_{5,\text{EF}}^0$ and $J_{5,\text{EF}}^z$ will depend on $G_A^Z(Q^2)$ and $G_T^Z(Q^2)$ respectively, in comparison with those BF ones (\ref{3DBF-distributions}). Moreover, we note that both $J_{5,\text{EF}}^0$ and $J_{5,\text{EF}}^z$ in the generic EF are free from the Wigner rotation (\ref{covariant-Lorentz-Transf}) while $\uvec J_{5,\text{EF}}^\perp$ suffers from it (see, e.g., the $\cos\theta$ and $\sin\theta$ factors in $\uvec J_{5,\text{EF}}^\perp$), since the Wigner rotation (\ref{covariant-Lorentz-Transf}) mixes $(\uvec\sigma_\perp)_{s' s}$ while leaves $(\sigma_z)_{s' s}$ and $(\uvec\Delta_\perp \cdot \uvec\sigma_\perp)_{s' s}$ unchanged~\cite{Chen:2023dxp}.

\begin{figure}[t!]
	\centering
	{\includegraphics[angle=0,scale=0.53]{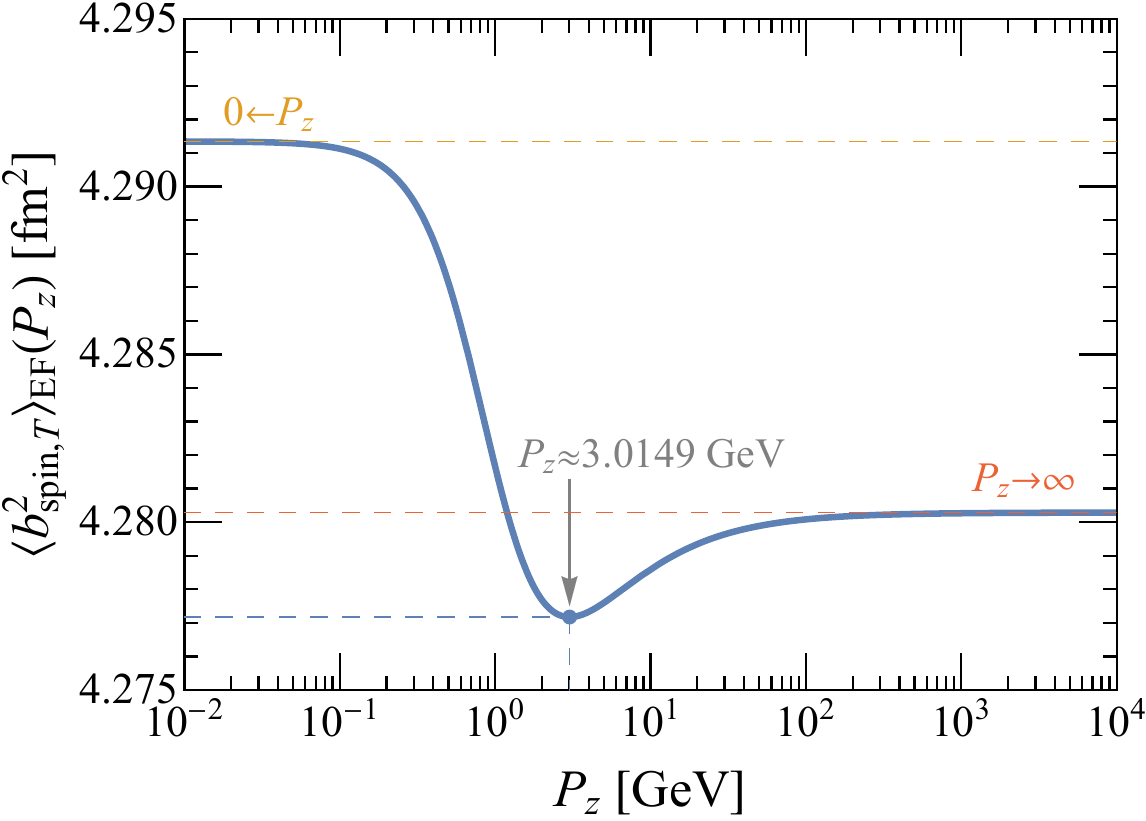}}
	\caption{The mean-square transverse spin radius $\langle b_{\text{spin},T}^2 \rangle_\text{EF}(P_z)$ of the proton as a function of $P_z$, using proton's weak-neutral axial-vector FFs $G_A^Z(0)$ and $G_P^Z(0)$ given in Appendix~\ref{Appendix-A}. The minimum of $\langle b_{\text{spin},T}^2 \rangle_\text{EF}(P_z)$ is located at $P_z \simeq 3.2132\,M_p \approx 3.0149~\text{GeV}$.}
	\label{Fig_2DEFNC_MSSpinT}
\end{figure}

\subsection{EF mean-square transverse radii}
\label{EF: Mean-square transverse radii}

Following Refs.~\cite{Chen:2023dxp,Chen:2024oxx}, we then rederive the mean-square transverse axial and spin radii using the EF distributions (\ref{2DEF-distributions}). Although $G_T^Z$ enters both $J_{5,\text{EF}}^0$ and $J_{5,\text{EF}}^z$ (\ref{2DEF-distributions}), we obtain exactly the same mean-square transverse radii as Ref.~\cite{Chen:2024oxx}
\begin{equation}
	\begin{aligned}\label{EF-MS-transverse-radii}
		\langle b_A^2 \rangle_\text{EF}(P_z)
		&= \frac{1}{2E_P^2} + \frac{2}{3} R_A^2,\\
		\langle b_{\text{spin},L}^2 \rangle_\text{EF}(P_z)
		&= \frac{2}{3} R_A^2,
	\end{aligned}
\end{equation}
with $E_P = \sqrt{M^2 + P_z^2}$. Likewise, since $G_T^Z(Q^2)$ does not enter $\uvec J_{5,\text{EF}}^\perp$, see Eq.~(\ref{2DEF-distributions}), we expect the same result for the mean-square transverse spin radius $\langle b_{\text{spin},T}^2 \rangle_\text{EF}(P_z)$ as Ref.~\cite{Chen:2024oxx}. From these results, we conclude that the second-class current contribution associated with $G_T^Z(Q^2)$, although explicitly included in our calculations, does not contribute to the mean-square transverse axial and spin radii in the generic EF.

Using the same expression of $\langle b_{\text{spin},T}^2 \rangle_\text{EF}(P_z)$ as Refs.~\cite{Chen:2024oxx,Chen:2024ksq} and the weak-neutral axial-vector FFs given in Appendix~\ref{Appendix-A}, we show in Fig.~\ref{Fig_2DEFNC_MSSpinT} the mean-square transverse spin radius $\langle b_{\text{spin},T}^2 \rangle_\text{EF}(P_z)$ for the proton as a function of $P_z$. In particular, we find that the minimum of $\langle b_{\text{spin},T}^2 \rangle_\text{EF}(P_z)$ is located at $P_{z} \simeq 3.2132~M_p \approx 3.0149~\text{GeV}$, while the maximum of $\langle b_{\text{spin},T}^2 \rangle_\text{EF}(P_z)$ is located at $P_z =0$. In the IMF limit $P_z \to \infty$, the value of $\langle b_{\text{spin},T}^2 \rangle_\text{EF}(\infty)$ is entirely contributed by the $P_z$-independent terms, see the Eq.~(52) of Ref.~\cite{Chen:2024ksq}, lying in between $\langle b_{\text{spin},T}^2 \rangle_\text{EF}(0)$ and $\langle b_{\text{spin},T}^2 \rangle_\text{EF}(P_z\simeq 3.2132~M_p)$.

\section{Light-front distributions}
\label{sec:Light-front distributions}

For completeness, we finally study the relativistic full weak-neutral axial-vector four-current distributions using the LF formalism~\cite{Brodsky:1997de}, where LF distributions in some cases can be interpreted as \emph{strict} probabilistic densities~\cite{Burkardt:2002hr,Miller:2007uy,Carlson:2007xd,Alexandrou:2008bn,Alexandrou:2009hs,Gorchtein:2009qq,Carlson:2009ovh,Miller:2010nz,Miller:2018ybm,Freese:2023jcp,Freese:2023abr} since the symmetry group associated with the transverse LF plane is the Galilean subgroup singled out from the Lorentz group~\cite{Susskind:1967rg,Kogut:1969xa}.

The $x^+$-independent symmetric LF frame~\cite{Burkardt:2002hr,Miller:2010nz,Lorce:2017wkb} is specified by the conditions\footnote{One can relax the condition $\uvec P_\perp=\uvec 0_\perp$, provided that LF distributions are restricted to $x^+=0$~\cite{Freese:2021czn,Freese:2022fat}.}: $\uvec P_\perp=\uvec 0_\perp$ and $\Delta^+=0$, which ensure that the LF energy transfer $\Delta^-=(\uvec P_\perp \cdot \uvec\Delta_\perp - P^- \Delta^+)/P^+$ vanishes automatically. Following Refs.~\cite{Lorce:2020onh,Chen:2023dxp,Chen:2024oxx}, the weak-neutral axial-vector four-current distributions in the symmetric LF frame are defined as
\begin{equation}\label{LF-def}
	\begin{aligned}
		J_{5,\text{LF}}^\mu(\uvec b_\perp;P^+)
		&\equiv \int \frac{\ud^2 \Delta_\perp}{(2\pi)^2}\, e^{-i\uvec\Delta_\perp \cdot \uvec b_\perp }\, \frac{ {_\text{LF}}\langle p',\lambda'| \hat{j}_5^\mu(0)|p,\lambda\rangle_\text{LF} }{2P^+} \bigg|_{\Delta^+=|\uvec P_\perp|=0},
	\end{aligned}
\end{equation}
where $P^+$ is treated as an independent variable in the LF formalism, and LF helicity states $|p,\lambda\rangle_\text{LF}$ are covariantly normalized as ${_\text{LF}}\langle p',\lambda'|p,\lambda\rangle_\text{LF}
= 2p^+ (2\pi)^3 \delta(p^{\prime\, +} - p^+) \delta^{(2)}( \uvec p_\perp^\prime -\uvec p_\perp ) \delta_{\lambda' \lambda}$. Besides, $|p,\lambda\rangle_\text{LF}$ can be related to the canonical spin states $|p, s \rangle$ via the Melosh rotation~\cite{Melosh:1974cu}
\begin{equation}
	\begin{aligned}\label{Melosh-rotation}
		|p,\lambda \rangle_\text{LF} = \sum_s \frac{ |p, s \rangle\langle p,s| }{ \langle p,s|p,s\rangle } |p,\lambda\rangle_\text{LF} = \sum_s |p,s \rangle \mathcal M_{s\lambda}^{(j)}(p),\qquad
		\mathcal M_{s\lambda}^{(j)}(p)\equiv \frac{ \langle p,s|p,\lambda\rangle_\text{LF} }{ \langle p,s|p,s\rangle },
	\end{aligned}
\end{equation}
where $\mathcal M^{(j)}(p)$ denotes the unitary Melosh rotation matrix for spin-$j$ systems. 

In the spin-$\frac{1}{2}$ case, the generic $2\times 2$ unitary Melosh rotation matrix $\mathcal M^{(1/2)}$ is explicitly given by~\cite{Lorce:2011zta,Chen:2022smg,Chen:2023dxp}
\begin{equation}
	\begin{aligned}\label{Melosh-rotation-M}
		\mathcal M^{(1/2)}(p)
		&= \begin{pmatrix}
			\cos\frac{\theta_M}{2} & -e^{-i\phi_p}\sin\frac{\theta_M}{2}\\
			e^{i\phi_p } \sin\frac{\theta_M}{2} & \cos\frac{\theta_M }{2}
		\end{pmatrix},
	\end{aligned}
\end{equation}
with
\begin{equation}
	\begin{aligned}\label{Melosh-angles}
		\cos\theta_M
		&= \frac{ (p^0 + p_z + M)^2 - |\uvec p_\perp|^2 }{ (p^0+p_z+m)^2 + \uvec p_\perp^2 },\qquad
		\sin\theta_M
		= -\frac{ 2(p^0 + p_z + M) |\uvec p_\perp| }{ (p^0+p_z+m)^2 + \uvec p_\perp^2 },
	\end{aligned}
\end{equation}
where $\uvec p_\perp = |\uvec p_\perp|(\cos\phi_p, \sin\phi_p)$, and $\theta_M$ is the \emph{Melosh rotation angle}. It is easy to verify that the condition $\cos^2\theta_M + \sin^2\theta_M=1$ is indeed automatically guaranteed. In the limit $p_z \to \infty$, the Melosh rotation matrix $\mathcal M^{(1/2)}$ becomes a $2\times 2$ identity matrix, and therefore the canonical spin polarization states $|s\rangle$ coincide with the LF helicity states $|\lambda \rangle_\text{LF}$ in the IMF limit, namely $\lim_{ p_z \to \infty}\, |s\rangle = |\lambda \rangle_\text{LF}$. 

We note that although the Melosh rotation matrix $\mathcal M^{(j)}$ becomes an trivial identity matrix in the IMF limit, it does not necessarily mean that LF distributions are completely free from relativistic artifacts caused by Melosh rotations~\cite{Lorce:2011zta,Lorce:2020onh,Lorce:2022jyi,Chen:2022smg,Chen:2023dxp}. For example, the change of polarization basis (\ref{Melosh-rotation}) modifies the normalization of four-momentum
eigenstates (i.e. $P^0 \to P^+$), which usually brings the explicit $P^+$-dependence for transverse LF distributions, e.g. $\uvec J_{5,\text{LF}}^\perp(\uvec b_\perp;P^+)$ in Eq.~(\ref{2DLF-distributions}) and $\uvec J_\text{LF}^\perp(\uvec b_\perp;P^+)$ in Ref.~\cite{Chen:2022smg}. Typically when a LF distribution is completely free from $P^+$, it should also free from relativistic artifacts caused by Melosh rotations and therefore assumes physically clear probabilistic interpretation~\cite{Soper:1976jc,Burkardt:2002hr,Miller:2007uy,Carlson:2007xd,Carlson:2009ovh,Miller:2010nz,Miller:2018ybm}, e.g. the LF electric charge distribution $J_\text{LF}^+$~\cite{Chen:2022smg} and the LF axial charge distribution $J_{5,\text{LF}}^+$ (\ref{2DLF-distributions}). From the Melosh rotation perspective, this also explains (to some extent) why the LF components $O^+$ and $O^-$ are usually regarded as the ``good'' and ``bad'' components in the literature, respectively. In this sense, we believe that the pictures provided by those LF densities that depend explicitly on $P^+$ can \emph{not} be considered as realistic representations of the system on the average at rest~\cite{Chen:2023dxp}.

\subsection{LF weak-neutral axial-vector four-current distributions}
\label{LF: Weak-neutral axial-vector four-current distributions}

By analogy with Eq.~(\ref{MatrElem-Axial-Z}), matrix elements of the axial-vector four-current operator in terms of LF helicity states $|p,\lambda \rangle_\text{LF}$ for a general spin-$\frac{1}{2}$ hadron are parametrized as
\begin{equation}
	\begin{aligned}\label{MatrElem-Axial-Z-LF}
		{_\text{LF}}\langle p',\lambda'| \hat{j}_5^\mu(0)|p,\lambda\rangle_\text{LF}
		&= \bar{u}_\text{LF}(p',\lambda')\left[ \gamma^\mu G_A^Z + \frac{\Delta^{\mu}}{2 M}G_P^Z - \frac{\sigma^{\mu\nu}\Delta_\nu }{2M} G_T^Z \right] \gamma^5 u_\text{LF}(p,\lambda),
	\end{aligned}
\end{equation}
where $u_\text{LF}(p,\lambda)$ denotes the LF helicity Dirac spinor, and the explicit $Q^2$-dependence of these axial-vector FFs $G_A^Z(Q^2)$, $G_P^Z(Q^2)$, and $G_T^Z(Q^2)$ for clarity is omitted.

Evaluating directly the matrix elements (\ref{MatrElem-Axial-Z-LF}) in the symmetric LF frame in terms of LF helicity Dirac spinors leads to~\cite{Chen:2024oxx}
\begin{equation}
	\begin{aligned}\label{LF-amplitudes}
		\mathcal A_\text{LF}^+
		&= 2P^+\left[  \frac{(i\uvec\Delta_\perp \cdot \uvec\sigma_\perp)_{\lambda' \lambda} }{2M}  G_T^Z(\uvec\Delta_\perp^2) + (\sigma_z)_{\lambda' \lambda} G_A^Z(\uvec\Delta_\perp^2) \right],\\
		\mathcal A_\text{LF}^-
		&= 2P^- \left[ \frac{(i\uvec\Delta_\perp \cdot \uvec\sigma_\perp)_{\lambda' \lambda} }{2M}  G_T^Z(\uvec\Delta_\perp^2) - (\sigma_z)_{\lambda' \lambda} G_A^Z(\uvec\Delta_\perp^2) \right],\\
		\uvec{\mathcal A}_{\text{EF}}^\perp
		&= 2M \bigg\{ \left[ (\uvec\sigma_\perp)_{\lambda' \lambda} +  \frac{(\uvec e_z \times i\uvec\Delta_\perp)_\perp }{2M} \delta_{ \lambda' \lambda } \right]G_A^Z(\uvec\Delta_\perp^2) - \frac{\uvec\Delta_\perp \left(\uvec\Delta_\perp \cdot \uvec\sigma_\perp \right)_{\lambda'\lambda} }{4M^2 } G_P^Z(\uvec\Delta_\perp^2) \bigg\},
	\end{aligned}
\end{equation}
with $\mathcal A_\text{LF}^\mu \equiv {_\text{LF}}\langle p',\lambda'|\hat j_{5}^{\mu}(0)|p,\lambda\rangle_\text{LF}$. We observe that $\uvec{\mathcal A}_{\text{EF}}^\perp$ is free from the second-class current contribution associated with $G_T^Z(Q^2)$, while $\mathcal A_\text{LF}^+$ and $\mathcal A_\text{LF}^-$ do receive contributions from that, similar as the case of the EF amplitudes (\ref{EF-amplitudes}).

Inserting the LF amplitudes (\ref{LF-amplitudes}) into Eq.~(\ref{LF-def}) leads to the following LF weak-neutral axial-vector four-current distributions~\cite{Chen:2024oxx,Diehl:2005jf}:
\begin{equation}
	\begin{aligned}\label{2DLF-distributions}
		J_{5,\text{LF}}^+(\uvec b_\perp;P^+)
		&= \int \frac{\ud^2 \Delta_\perp}{(2\pi)^2}\, e^{-i\uvec\Delta_\perp \cdot \uvec b_\perp }\, \left[ \frac{ (i\uvec\Delta_\perp \cdot \uvec\sigma_\perp)_{\lambda' \lambda} }{2M} G_T^Z(\uvec\Delta_\perp^2) + (\sigma_z)_{\lambda' \lambda} G_A^Z(\uvec\Delta_\perp^2) \right],\\
		J_{5,\text{LF}}^-(\uvec b_\perp;P^+)
		&= \int \frac{\ud^2 \Delta_\perp}{(2\pi)^2}\, e^{-i\uvec\Delta_\perp \cdot \uvec b_\perp }\, \frac{P^-}{P^+} \left[ \frac{ (i\uvec\Delta_\perp \cdot \uvec\sigma_\perp)_{\lambda' \lambda} }{2M}  G_T^Z(\uvec\Delta_\perp^2) - (\sigma_z)_{\lambda' \lambda} G_A^Z(\uvec\Delta_\perp^2) \right] ,\\
		\uvec J_{5,\text{LF}}^\perp(\uvec b_\perp;P^+)
		&= \int\frac{\ud^2\Delta_\perp }{(2\pi)^2 }\, e^{-i\uvec\Delta_\perp \cdot \uvec b_\perp }\, \frac{M}{P^+} \bigg\{  \left[ (\uvec\sigma_\perp)_{\lambda' \lambda} +  \frac{(\uvec e_z \times i\uvec\Delta_\perp)_\perp }{2M} \delta_{ \lambda' \lambda } \right]G_A^Z(\uvec\Delta_\perp^2)\\
		&\qquad - \frac{\uvec\Delta_\perp \left(\uvec\Delta_\perp \cdot \uvec\sigma_\perp \right)_{\lambda'\lambda} }{ 4M^2 } G_P^Z(\uvec\Delta_\perp^2)  \bigg\}.
	\end{aligned}
\end{equation}
We remind that $J_{5,\text{LF}}^+$ is also called the LF helicity distribution~\cite{Burkardt:2002hr,Diehl:2005jf}, and $\uvec S_\text{LF}^\perp(\uvec b_\perp;P^+) = \uvec J_{5,\text{LF}}^\perp(\uvec b_\perp;P^+)/2$ is the transverse LF spin distribution~\cite{Lorce:2017wkb,Lorce:2018egm,Lorce:2022cle,Chen:2024oxx}.

According to Eqs.~(\ref{2DLF-distributions}) and (\ref{2DEF-distributions}), it is clear that
\begin{equation}\label{EF-IMF-LF-equal}
	J_{5,\text{LF}}^+(\uvec b_\perp;P^+) = J_{5,\text{EF}}^0(\uvec b_\perp;\infty)
	= J_{5,\text{EF}}^z(\uvec b_\perp;\infty),
\end{equation}
where we used the fact that canonical spin polarizations coincide with LF helicities in the IMF limit. Above result (\ref{EF-IMF-LF-equal}) is quite reminiscent of the similar results in the electromagnetic case~\cite{Chen:2022smg,Chen:2023dxp}. It is therefore clear that for a longitudinally polarized (i.e. $\hvec s = \uvec e_z$) proton, $J_{5,\text{LF}}^+(b;P^+)$ and $2\pi b \cdot J_{5,\text{LF}}^+(b;P^+)$ will be exactly the same as the $P_z = \infty$ case of $J_{5,\text{EF}}^0(b;P_z) $ and $2\pi b \cdot J_{5,\text{EF}}^0(b;P_z) $ in Fig.~\ref{Fig_2DEFNC_J0Sz}, respectively. Similarly for a transversely polarized (i.e. $\hvec s = \uvec e_x$) proton, $J_{5,\text{LF}}^+(\uvec b_\perp;P^+)$ and $2\pi b \cdot J_{5,\text{LF}}^+(\uvec b_\perp;P^+)$ will be exactly the same as the $P_z = \infty$ case of $J_{5,\text{EF}}^z(\uvec b_\perp;P_z) $ and $2\pi b \cdot J_{5,\text{EF}}^z(\uvec b_\perp;P_z) $ in Fig.~\ref{Fig_2DEFNC_JzSx}, respectively. 

\begin{figure}[t!]
	\centering
	{\includegraphics[angle=0,scale=0.52]{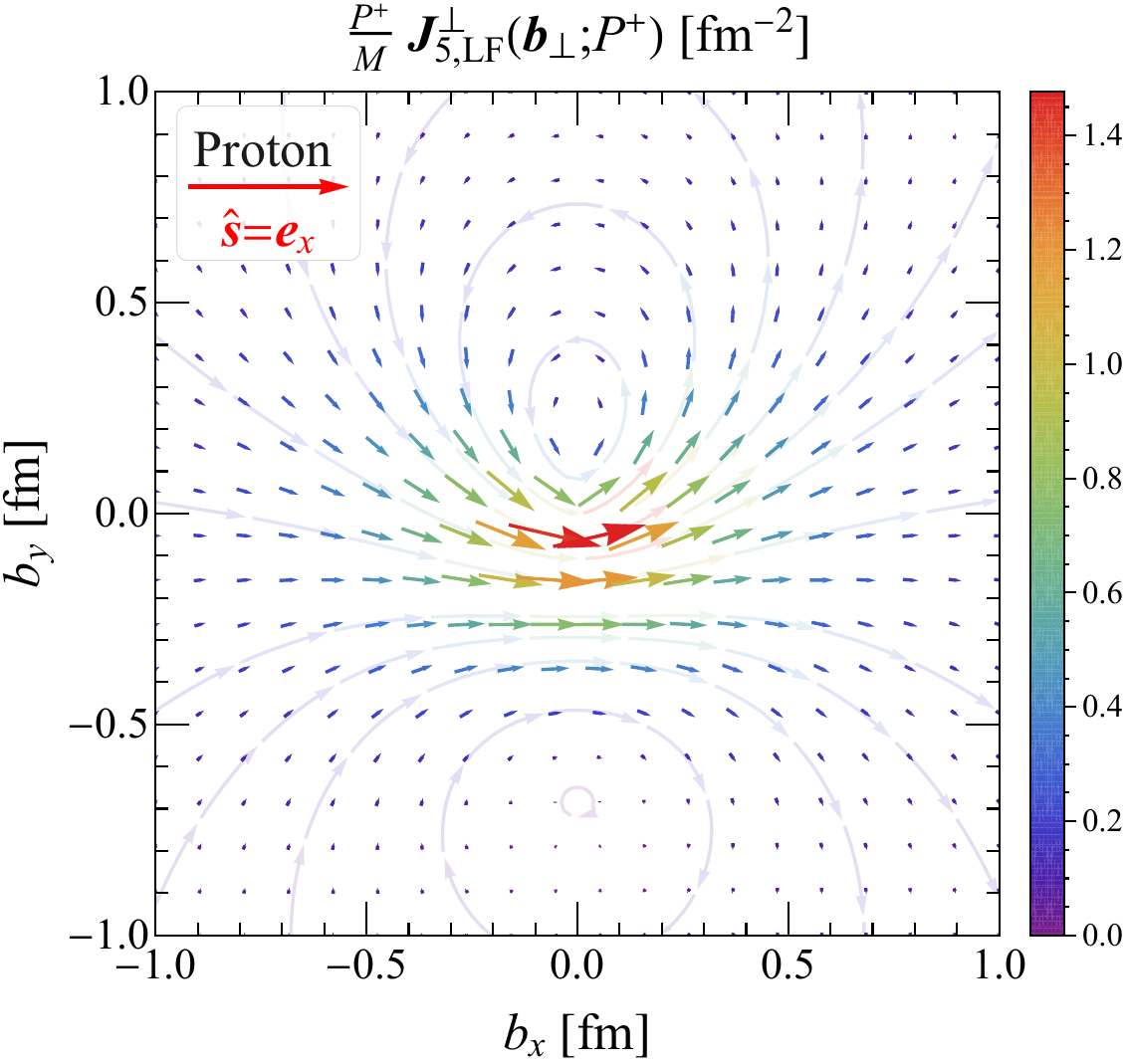}}
	\caption{The (scaled) transverse LF weak-neutral axial-vector current distribution $\tfrac{P^+}{M}\uvec J_{5,\text{LF}}^\perp(\uvec b_\perp;P^+)$ in the transverse plane inside a transversely polarized (i.e. $\hvec s = \uvec e_x$) proton, using proton's weak-neutral axial-vector FFs $G_A^Z(Q^2)$ and $G_P^Z(Q^2)$ given in Appendix~\ref{Appendix-A}.}
	\label{Fig_2DLFNC_JvT}
\end{figure}

In Fig.~\ref{Fig_2DLFNC_JvT}, we illustrate the (scaled) transverse LF axial-vector current distribution $\tfrac{P^+}{M}\uvec J_{5,\text{LF}}^\perp(\uvec b_\perp;P^+)$ in the transverse plane inside a transversely polarized proton, using proton's weak-neutral axial-vector FFs given in Appendix~\ref{Appendix-A}. Similarly as Fig.~\ref{Fig_2DEFNC_JvT}, the distribution $\tfrac{P^+}{M}\uvec J_{5,\text{LF}}^\perp$ in the transverse plane is not mirror symmetric with respect to the $x$-axis but is still mirror antisymmetric with respect to the $y$-axis, and it also exhibits the toroidal mode~\cite{Repko:2012rj,vonNeumann-Cosel:2023aep} in the weak sector similar as Fig.~\ref{Fig_2DEFNC_JvT}. This is consistent with our expectation, since $\uvec J_{5,\text{LF}}^\perp$ also contains monopole, dipole and quadrupole contributions similar as $\uvec J_{5,\text{EF}}^\perp$.

\subsection{LF amplitudes via proper IMF limit of EF amplitudes}
\label{LF: LF amplitudes via proper IMF limit of EF amplitudes}

Following Refs.~\cite{Chen:2022smg,Lorce:2024ipy,Lorce:2024ewq}, we can explicitly reproduce the genuine LF amplitudes (\ref{LF-amplitudes}) by making use of the EF amplitudes (\ref{EF-amplitudes}) in the \emph{proper} IMF limit [i.e. expand the amplitudes in $1/P^+$ at large $P^+$, and keep only the leading term]\footnote{More specifically, we first switch variables from $\{P^0, P_z \}$ to $\{P^+, P^-\}$ in the constructed EF amplitudes, and then expand the amplitudes in $1/P^+$ at large $P^+$, and finally keep only the leading term.}. Starting from Eq.~(\ref{EF-amplitudes}), we can first construct the following amplitudes $\mathcal A^\pm_\text{EF}$ in the generic EF:
\begin{equation}\label{construced-EF-amplitudes}
	\begin{aligned}
		\langle p',s'| \hat j_5^+(0) |p, s\rangle
		&= \frac{ \mathcal A_\text{EF}^0 + \mathcal A_\text{EF}^z }{ \sqrt{2} } = 2P^+ \left[ \frac{ (i\uvec\Delta_\perp \cdot \uvec\sigma_\perp )_{s's} }{2M}  G_T^Z(\uvec\Delta_\perp^2) + (\sigma_z)_{s's} G_A^Z(\uvec\Delta_\perp^2) \right],\\
		\langle p',s'| \hat j_5^-(0) |p, s\rangle
		&= \frac{ \mathcal A_\text{EF}^0 - \mathcal A_\text{EF}^z }{ \sqrt{2} } = 2P^-\left[ \frac{ (i\uvec\Delta_\perp \cdot \uvec\sigma_\perp )_{s's} }{2M}  G_T^Z(\uvec\Delta_\perp^2) - (\sigma_z)_{s' s} G_A^Z(\uvec\Delta_\perp^2) \right],
	\end{aligned}
\end{equation}
with $\hat j_5^\pm \equiv (\hat j_5^0 \pm \hat j_5^z)/\sqrt{2}$. Clearly, above amplitudes $\langle p',s'| \hat j_5^\pm (0) |p, s\rangle$ are not proper LF amplitudes since they are still defined in terms of the canonical (or instant-form) polarization states $|s \rangle$ rather than the LF helicity states $|\lambda \rangle_\text{LF}$~\cite{Chen:2022smg,Lorce:2024ipy,Lorce:2024ewq}. 

We are now ready to take the proper IMF limit of $\langle p',s'| \hat j_5^\pm(0) |p, s\rangle$ and $\langle p',s'| \hat{\uvec j}_5^\perp(0) |p, s\rangle$. It then follows that the constructed amplitudes are given by
\begin{equation}\label{construced-IMF-amplitudes}
	\begin{aligned}
		\langle p',s'| \hat j_5^+(0) |p, s\rangle \Big|_{\text{proper IMF}}
		&= 2P^+ \left[ \frac{ (i\uvec\Delta_\perp \cdot \uvec\sigma_\perp )_{s's}}{2M}  G_T^Z(\uvec\Delta_\perp^2) + (\sigma_z)_{s's} G_A^Z(\uvec\Delta_\perp^2) \right],\\
		\langle p',s'| \hat j_5^-(0) |p, s\rangle\Big|_{\text{proper IMF}}
		&= 2P^-\left[ \frac{ (i\uvec\Delta_\perp \cdot \uvec\sigma_\perp )_{s' s} }{2M}  G_T^Z(\uvec\Delta_\perp^2) - (\sigma_z)_{s' s} G_A^Z(\uvec\Delta_\perp^2) \right],\\
		\langle p',s'| \hat{\uvec j}_5^\perp(0) |p, s\rangle \Big|_{\text{proper IMF}}
		&= 2M\left[ (\uvec\sigma_\perp)_{s' s} + \frac{(\uvec e_z \times i\uvec\Delta_\perp)_\perp  }{2M} \delta_{s' s}  \right]G_A^Z(\uvec\Delta_\perp^2)\\
		&\quad - \frac{\uvec\Delta_\perp \left(\uvec\Delta_\perp \cdot \uvec\sigma_\perp \right)_{s' s} }{2M} G_P^Z(\uvec\Delta_\perp^2),
	\end{aligned}
\end{equation}
We note that in the IMF limit (i.e., $P_z \to \infty$), similarly as in the electromagnetic case~\cite{Chen:2022smg}, the amplitude $\langle p',s'| \hat j_5^+(0) |p, s\rangle$ will be enhanced while the amplitude $\langle p',s'| \hat j_5^-(0) |p, s\rangle$ will be suppressed, owing to the associated global factors $P^+$ and $P^-$, respectively. Using the fact that $\lim_{P_z \to \infty }|s \rangle = |\lambda \rangle_\text{LF}$ since $\mathcal M^{(1/2)}(p)=\mathcal M^{\dag (1/2)}(p') = \mathds1_{2\times 2}$ at $P_z \to \infty$, we can simply apply the replacements $s \to \lambda$ and $s' \to \lambda'$ to the constructed amplitudes (\ref{construced-IMF-amplitudes}). In doing so, we indeed explicitly reproduce the genuine LF amplitudes (\ref{LF-amplitudes}). Following this procedure, one can also explicitly reproduce the LF amplitudes (\ref{LF-amplitudes}) by using the EF amplitudes (\ref{EF-amplitudes-2}) in terms of Wigner rotation angle $\theta$, with the help that~\cite{Chen:2022smg}\footnote{We note that there is a typo in the Eq.~(41) of Ref.~\cite{Chen:2022smg} that ``$\lim_{P_z \to \infty}\, \tan\theta = -1/\sqrt{\tau}$'' should be corrected as ``$\lim_{P_z \to \infty}\, \tan\theta = -\sqrt{\tau}$'', see Eq.~(\ref{Wigner-angles-IMF}).}
\begin{equation}\label{Wigner-angles-IMF}
	\lim_{P_z \to \infty }\, \cos\theta 
	= \frac{1}{ \sqrt{1+\tau} },\qquad \lim_{P_z \to \infty }\, \sin\theta = -\frac{ \sqrt{\tau} }{ \sqrt{1+\tau} }.
\end{equation}
We remind that similar procedure has been used for the cross check of the LF polarization and magnetization amplitudes in Ref.~\cite{Chen:2023dxp}.

Explicit demonstrations of LF amplitudes via the proper IMF limit of corresponding EF amplitudes in electromagnetic four-current~\cite{Chen:2022smg,Lorce:2024ipy,Lorce:2024ewq}, polarization-magnetization tensor~\cite{Chen:2023dxp}, and axial-vector four-current (\ref{construced-EF-amplitudes}-\ref{Wigner-angles-IMF}) cases inspire us to propose the following conjecture:
\begin{framed}
	\begin{conjecture}
		Any light-front (LF) amplitudes for well-defined LF distributions in principle can be explicitly reproduced from the corresponding elastic frame (EF) amplitudes in the proper infinite-momentum frame (IMF) limit.
	\end{conjecture}
\end{framed}

As a reward, the origins of distortions appearing in LF distributions (relative to the BF ones) can be understood more easily and intuitively using the covariant Lorentz transformation and the above conjecture, with the help of the Melosh rotation (\ref{Melosh-rotation}). We should note that to obtain LF amplitudes one does not necessarily need to first perform the covariant Lorentz transformation and then reproduce the LF amplitudes using above conjecture, since one can always straightforwardly obtain the LF amplitudes by directly evaluating LF helicity Dirac bilinears. However, it is usually \emph{not} so easy and intuitive to understand the distortions appearing in LF distributions via the direct evaluation of LF helicity Dirac bilinears. Our procedure provides a legitimate but more easy and intuitive way to understand the origins of distortions appearing in LF distribution (relative to the BF ones).

Hence, we can classify more easily the origins of distortions appearing in LF distributions (relative to the BF ones) into three key sources. The first key source of distortions in LF distributions is due to the peculiar LF perspective for defining the $O^+$ and $O^-$ components~\cite{Chen:2022smg,Chen:2023dxp}. In the cases of electromagnetic and axial-vector four-current amplitudes, the expressions of LF amplitudes $\widetilde O_\text{LF}^\pm$ involve $\gamma(1\pm \beta) \, \widetilde O_B^\pm$ under a longitudinal Lorentz boost, where $\widetilde O_B^\pm \equiv (\widetilde O_B^0 \pm \widetilde O_B^3)/\sqrt{2}$ stand for the constructed BF amplitudes. The $\gamma(1\pm \beta)$ factors will result in the overall $P^\pm$ factors in $\widetilde O_\text{LF}^\pm$, respectively. According to the generic definition of LF distributions $O_\text{LF}^\mu(\uvec b_\perp;P^+)$~\cite{Chen:2022smg,Chen:2023dxp,Chen:2024oxx}, the $P^+$ factor in $\widetilde O_\text{LF}^+$ will be canceled out by the same $P^+$ factor in the definition of $O_\text{LF}^\mu(\uvec b_\perp;P^+)$ arising due to the normalization of LF four-momentum eigenstates. Therefore, LF distributions $O_\text{LF}^+(\uvec b_\perp;P^+)$ are usually $P^+$-independent, allowing therefore physically clear probabilistic interpretation~\cite{Soper:1976jc,Burkardt:2002hr,Miller:2010nz}. However, the $P^-$ factor in $\widetilde O_\text{LF}^-$ can not be canceled out and results in an $Q$-dependent factor $P^-/P^+ = M^2(1+\tau)/[2(P^+)^2]$ within the Fourier transform, which together with $\widetilde O_B^-$ usually causes strange distortions and distributions. As a result, LF distributions $O_\text{LF}^-(\uvec b_\perp;P^+)$ are usually regarded as the ``bad'' components and are considered as complicated objects without clear physical interpretation~\cite{Chen:2022smg}. For example, physically clear $F_1(Q^2)$ and $F_2(Q^2)$ in $J_\text{LF}^+$ become physically unclear $G_1(Q^2)$ and $G_2(Q^2)$ in $J_\text{LF}^-$ in the electromagnetic case~\cite{Chen:2022smg}, and the sum of the two terms associated respectively with $G_T^Z(Q^2)$ and $G_A^Z(Q^2)$ in $J_{5,\text{LF}}^+$ (\ref{2DLF-distributions}) becomes the difference between these two terms in $J_{5,\text{LF}}^-$ (\ref{2DLF-distributions}).

The second key source of distortions originates from complicated Wigner rotations and the Lorentz mixing of temporal and longitudinal components under longitudinal Lorentz boosts according to the covariant Lorentz transformation for amplitudes from BF to the IMF [i.e. $\widetilde O_B^\mu \to \widetilde O_\text{IMF}^\mu$], see e.g. Eq.~(\ref{covariant-Lorentz-Transf}). This is the only source of physical distortions for EF and IMF distributions. Based on the conjecture above, one needs to take the proper IMF limit of EF amplitudes, which in general causes kinematical suppression (or Lorentz contraction) for terms in the EF amplitudes involving $P^0$ in the denominator while there is no $P_z$-dependent factor in the numerator, e.g. the term involving $G_A^Z(Q^2)/(P^0+M)$ in $\uvec{\mathcal A}_\text{EF}^\perp$~(\ref{EF-amplitudes}) is completely suppressed at $P_z \to \infty$.

The third key source of distortions originates from relativistic artefacts caused by Melosh rotations (\ref{Melosh-rotation}). The change of polarization basis modifies the normalization of four-moment eigenstates, which usually brings the explicit $P^+$-dependence for the transverse components of LF distributions, see e.g. $\uvec J_{5,\text{LF}}^\perp(\uvec b_\perp;P^+)$ in Eq.~(\ref{2DLF-distributions}) and $\uvec J_\text{LF}^\perp(\uvec b_\perp;P^+)$ in Ref.~\cite{Chen:2022smg}.

\subsection{LF mean-square transverse radii}
\label{LF: mean-square transverse radii}

Following Refs.~\cite{Chen:2024oxx,Diehl:2005jf}, we also rederive the mean-square transverse axial (or helicity) and spin radii using the LF distributions (\ref{2DLF-distributions}). We find that
\begin{equation}
	\begin{aligned}\label{LF-MS-transverse-radii}
		\langle b_A^2 \rangle_\text{LF}(P^+)
		&= \langle b_{\text{spin},L}^2 \rangle_\text{LF}(P^+) = \frac{2}{3} R_A^2,\\
		\langle b_{\text{spin},T}^2 \rangle_\text{LF}(P^+) 
		&= \frac{2}{3} R_A^2 + \frac{1}{2M^2} \frac{G_P^Z(0)}{G_A^Z(0) },
	\end{aligned}
\end{equation}
which do coincide with Ref.~\cite{Chen:2024oxx} although we have taken into account $G_T^Z(Q^2)$, see Eq.~(\ref{2DLF-distributions}). From above results (\ref{LF-MS-transverse-radii}), we conclude that the second-class current contribution associated with the induced pseudotensor FF $G_T^Z(Q^2)$, although explicitly included in our calculations, does not contribute in fact to the mean-square transverse axial and spin radii on the LF.

\section{Summary}
\label{sec:Summary}

In this paper, we extended our recent work~\cite{Chen:2024oxx} of the relativistic weak-neutral axial-vector four-current distributions inside a general spin-$\frac{1}{2}$ composite system, where the second-class current contribution associated with the weak-neutral induced pseudotensor form factor is newly included. To the best of our knowledge, this is the first time that the full weak-neutral axial-vector four-current distributions inside a general spin-$\frac{1}{2}$ hadron are systematically studied in terms of relativistic Breit frame, elastic frame and light-front distributions.

When the system is on the average at rest, we clearly demonstrated that the 3D weak-neutral axial charge distribution $J^0_{5,B}$ in the Breit frame is in fact related to the induced pseudotensor form factor $G_T^Z(Q^2)$ rather than the axial form factor $G_A^Z(Q^2)$. Besides, we found that $J^0_{5,B}$ is of parity-odd nature (\ref{3DBF-distributions}), due to which the definition of the standard mean-square axial (charge) radius~(\ref{3DBF-MS-axial-radius}) is in fact not well-defined. We also tested for the first time the Abel and its inverse transforms of axial charge distributions in the spin-$\frac{1}{2}$ case, through which we explicitly revealed the breakdown of Abel transformation for the connection in physics between 2D light-front and 3D Breit frame axial charge distributions, see Appendix~\ref{Appendix-B}. On the other hand, we found that the 3D axial-vector current distribution $\uvec J_{5,B}$ is free from the second-class current contribution, and is closely related to the physically meaningful 3D (intrinsic) spin distribution~\cite{Lorce:2017wkb,Lorce:2018egm,Lorce:2022cle,Chen:2024oxx}.

When the system is boosted, the situation gets more complicated since both the Wigner rotation and the Lorentz mixing effect will play the roles. We did observe clear frame-dependence of the axial charge distribution $J_{5,\text{EF}}^0$ (longitudinal axial-vector current distribution $J_{5,\text{EF}}^z$) in the generic elastic frame for a longitudinally (transversely) polarized target. The frame-dependence of $J_{5,\text{EF}}^0$ and $J_{5,\text{EF}}^z$ is solely due to the Lorentz mixing of temporal and longitudinal components of the axial-vector four-current amplitudes under longitudinal Lorentz boosts (\ref{covariant-Lorentz-Transf}), since they are both free from the Wigner rotation. On the contrary, the transverse axial-vector current distribution $\uvec J_{5,\text{EF}}^\perp$~(\ref{2DEF-distributions}) does not get mixed under longitudinal Lorentz boosts, but it suffers from the Wigner rotation (\ref{EF-amplitudes-2}).

We also studied the full weak-neutral axial-vector four-current distributions using the light-front formalism. We revisited the role played by Melosh rotations, and further proposed the conjecture based on our recent works~\cite{Chen:2022smg,Chen:2023dxp,Chen:2024oxx,Lorce:2024ipy,Lorce:2024ewq} that {\em any light-front amplitudes for well-defined light-front distributions in principle can be explicitly reproduced from the corresponding elastic frame amplitudes in the proper infinite-momentum frame limit}. As a reward, we can understand more easily and intuitively the origins of distortions appearing in light-front distributions (relative to the Breit frame ones). For completeness, we also rederived the mean-square transverse axial and spin radii in different frames. We showed in particular that the second-class current contribution, although explicitly included in our calculations, does not contribute in fact to the mean-square axial and spin radii.

To get a more intuitive picture of the weak-neutral axial-vector structure of a spin-$\frac{1}{2}$ hadron, we also numerically illustrated our results in the case of a proton, using proton's weak-neutral axial-vector form factors extracted from experimental data, see Appendix~\ref{Appendix-A}. It should be emphasize that our analytic formulas apply to any spin-$\frac{1}{2}$ hadrons (e.g. $\Lambda^0$, $\Sigma^0$, $\Xi^0$, etc.) and can be easily generalized to higher spin systems, as long as the corresponding weak-neutral axial-vector form factors are available.

\appendix
\section{Parametrization of nucleon weak-neutral axial-vector FFs}
\label{Appendix-A}

In the literature~\cite{Horstkotte:1981ne,Ahrens:1986xe,MiniBooNE:2010xqw,MiniBooNE:2013dds,Bernard:2001rs}, the nucleon weak-neutral $G_A^Z(Q^2)$ and weak-charged $G_A^W(Q^2)$ axial FFs are usually parametrized in terms of a standard dipole model \emph{ans\"atz}:
\begin{equation}
	\begin{aligned}\label{NC-Z-FF-GA}
		G_A^L(Q^2) = \frac{G_A^L(0) }{ \left( 1 + \frac{Q^2}{ (M_A^L)^2 } \right)^2 },
	\end{aligned}
\end{equation}
where $L=Z,\, W$ and $M_A^L$ is the corresponding (axial) dipole mass. We note that the weak-charged nucleon axial FF $G_A^W(Q^2)$ has been extracted from quasielastic (anti)neutrino-nucleon and (anti)neutrino-nuclei scattering data with $M_A^W \approx (1.026 \pm 0.021)\,\text{GeV}$~\cite{Bernard:2001rs}. The weak-charged axial charge (or coupling constant) $G_A^W(0) = (1.2754 \pm 0.0013)$~\cite{ParticleDataGroup:2024cfk} is very precisely determined in neutron beta decay reaction $n \to p e^- \bar\nu_e$.

According to Refs.~\cite{Weinberg:1972tu,Horstkotte:1981ne,Ahrens:1986xe,Garvey:1992cg,Garvey:1993sg,Pate:2003rk,Sufian:2018qtw}, the weak-neutral axial-vector FFs $G_X^Z(Q^2)$ for $X=A,P,T$ can be related to the corresponding weak-charged ones $G_X^W(Q^2)$ via
\begin{equation}
	\begin{aligned}\label{Relations-NC-CC}
		G_X^Z(Q^2)
		&= \sum_f g_A^f \, G_X^f(Q^2)\\
		&\simeq \frac{1}{2}\left[ G_X^{W}(Q^2) - G_X^s(Q^2) + G_X^c(Q^2) - G_X^b(Q^2) + G_X^t(Q^2) \right],
	\end{aligned}
\end{equation}
where $G_X^W\simeq G_X^u - G_X^d \equiv G_X^{(u-d)}$, and $G_X^f$ denotes the FF contribution from the $f$-flavor quark with $f=u,d,s,c,b,t$. We note that the axial-vector couplings of quarks to the $Z^0$ boson in the Standard Model are given by $g_A^{u,c,t}=\frac{1}{2}$ and $g_A^{d,s,b}=-\frac{1}{2}$, which explain the overall factor $\frac{1}{2}$ in Eq.~(\ref{Relations-NC-CC}). It should be noted that $G_A^Z(t=-Q^2)$ and $G_P^Z(t=-Q^2)$ can also be accessed via measurements of generalized parton distributions (GPDs) $\tilde H_f(x,\xi,t)$ and $\tilde E_f(x,\xi,t)$ at JLab (Hall A/B/C), LHC, HERA (HERMES, H1, ZEUS), NICA, EIC, EicC, etc. through the first-moment sum rules~\cite{Ji:1996ek,Diehl:2003ny,JeffersonLabHallA:2022pnx}
\begin{equation}
	\begin{aligned}\label{relation_to_GPDs}
		\begin{bmatrix}
			G_A^Z(t)\\
			G_P^Z(t)
		\end{bmatrix}
		= \sum_f g_A^f \int_{-1}^{1} \ud x
		\begin{bmatrix}
			\tilde H_f(x,\xi,t)\\
			\tilde E_f(x,\xi,t)
		\end{bmatrix}.
	\end{aligned}
\end{equation}
In general, the contributions from the heavy-flavor quarks (i.e. $c$, $b$, $t$) are very small for the nucleon and are therefore usually neglected in practical calculations. In this analysis, we also neglect the contributions from heavy-flavor quarks for the nucleon. 

Using directly the extracted experimental data of the proton weak-neutral axial FF $G_A^Z(Q^2)$ from Ref.~\cite{Sufian:2018qtw} based on recent MiniBooNE measurements~\cite{MiniBooNE:2010xqw,MiniBooNE:2013dds} and performing the standard dipole model (\ref{NC-Z-FF-GA}) fit to the data (denoted as ``Dipole fit''), we find
\begin{equation}\label{NC-Z-FF-MA}
	M_A^Z \approx (1.0500 \pm 0.0107)~\text{GeV},
\end{equation}
where $G_A^Z(0) = (0.65520 \pm 0.00465)$ is fixed using both world average experimental data of $G_A^W(0) = (1.2754 \pm 0.0013)$ from Particle Data Table~\cite{ParticleDataGroup:2024cfk} and the strange quark contribution to the nucleon spin $\Delta s \equiv G_A^s(0) \approx (-0.0350 \pm 0.0092)$ in the continuum limit and physical pion point from lattice QCD~\cite{Liang:2018pis}; see also Refs.~\cite{Pate:2003rk,Pate:2008va,Pate:2024acz} for recent analyses of $G_A^s(0)$.

\begin{figure}[t!]
	\centering
	{\,\includegraphics[angle=0,scale=0.39]{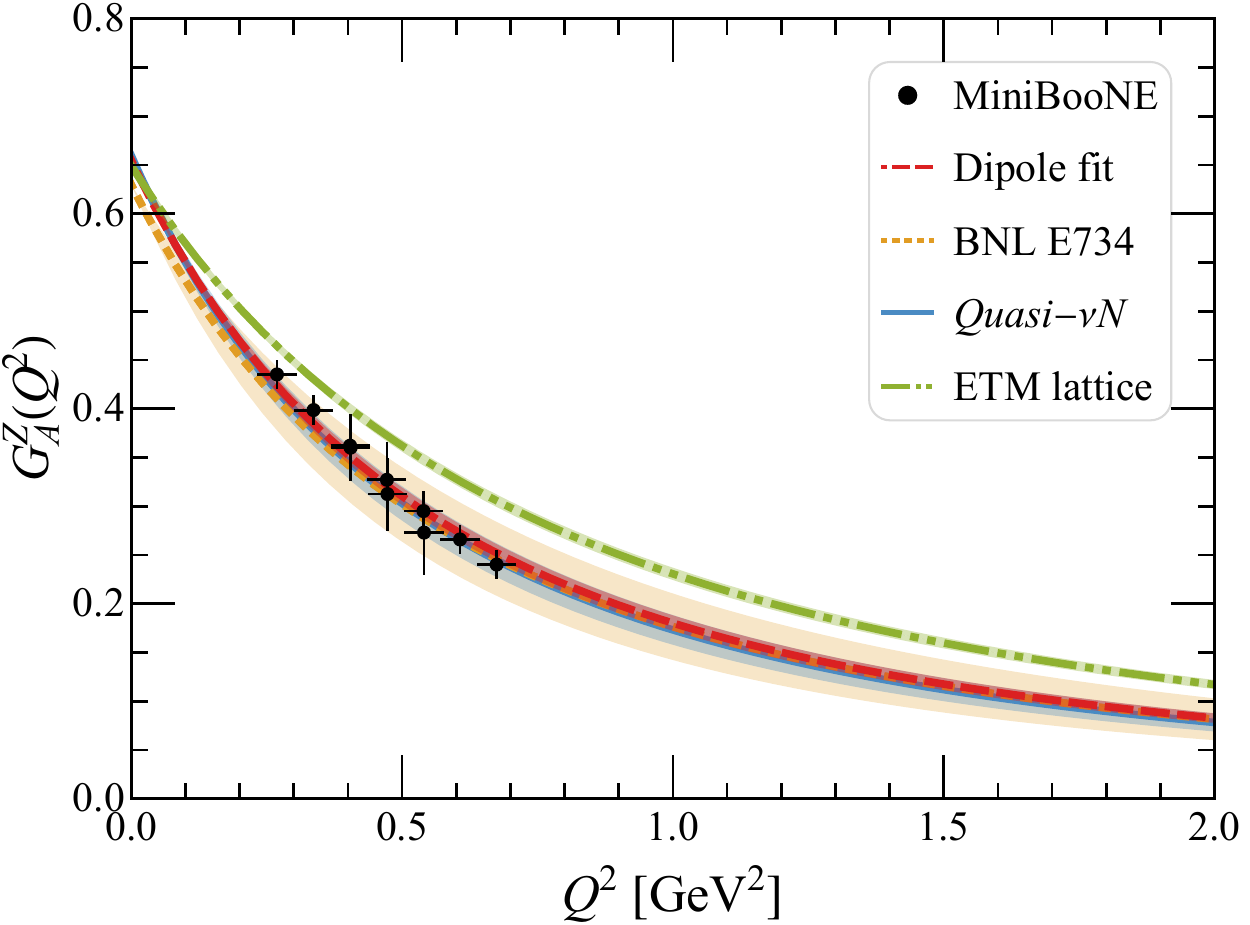}}
	{\includegraphics[angle=0,scale=0.396]{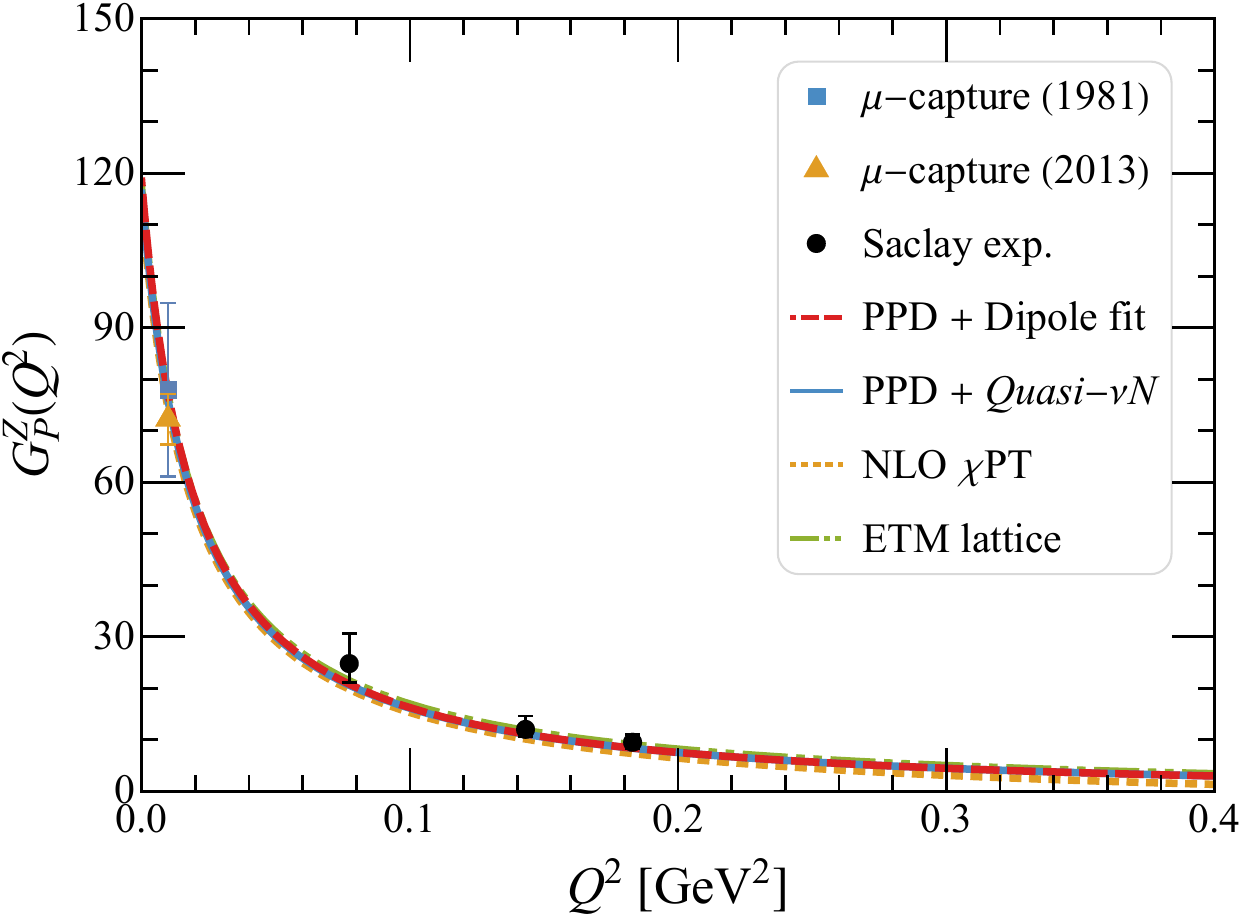}}
	\caption{Comparison of the nucleon weak-neutral axial $G_A^Z(Q^2)$ (upper panel) and induced pseudoscalar $G_P^Z(Q^2)$ (lower panel) FFs as a function of $Q^2$ using different methods, where confidence bands of $G_A^Z(Q^2)$ at $95\%$ confidence level are also shown. See texts for more details.}
	\label{Fig_weakNeutral_GAGP}
\end{figure}

Alternative to Eq.~(\ref{NC-Z-FF-GA}) with $M_A^Z$ given in Eq.~(\ref{NC-Z-FF-MA}), we can also obtain $G_A^Z(Q^2)$ in terms of $G_A^W(Q^2)$ and $G_A^s(Q^2)$ via Eq.~(\ref{Relations-NC-CC})\footnote{We note that the direct (anti)neutrino-nucleon elastic scattering data~\cite{MiniBooNE:2010xqw,MiniBooNE:2013dds,Sufian:2018qtw} of $G_A^Z(Q^2)$ in turn can help us to test whether the relation (\ref{Relations-NC-CC}) between $G_A^Z(Q^2)$ and $G_A^W(Q^2)$ is valid or not.}. For the strange quark contributions $G_A^s(Q^2)$ and $G_P^s(Q^2)$, $G_X^s(Q^2)$ for $X=A,P$ are also usually  parametrized in the literature in terms of the standard dipole model~\cite{Alexandrou:2021wzv}
\begin{equation}
	\begin{aligned}\label{NC-Z-FF-s-quark}
		G_X^s(Q^2) = \frac{ G_X^s(0) }{ \left( 1 + \frac{Q^2}{ (M_X^s)^2 } \right)^2 },
	\end{aligned}
\end{equation}
where the corresponding dipole parameters $G_A^s(0)=-0.044(8)$, $M_A^s=0.992(164)~\text{GeV}$, $G_P^s(0)=-1.325(406)$ and $M_P^s=0.609(89)~\text{GeV}$ can be found in Ref.~\cite{Alexandrou:2021wzv}. 

In the upper panel of Fig.~\ref{Fig_weakNeutral_GAGP}, we show the direct ``Dipole fit'' (red dot-dashed curve) result of $G_A^Z(Q^2)$ using the extracted MiniBooNE data~\cite{MiniBooNE:2010xqw,MiniBooNE:2013dds,Sufian:2018qtw} (circle markers) via Eq.~(\ref{NC-Z-FF-GA}) and the reconstructed $G_A^Z(Q^2)=[G_A^{W}(Q^2) - G_A^s(Q^2)]/2$ labeled by ``\emph{Quasi}-$\nu N$'' (blue solid curve) using Eq.~(\ref{Relations-NC-CC}) in terms of $G_A^W$ (\ref{NC-Z-FF-GA}) extracted from quasielastic (anti)neutrino scattering data and $G_A^s$ (\ref{NC-Z-FF-s-quark}), in comparison with $G_A^Z(Q^2)$ from the BNL E734 measurements~\cite{Horstkotte:1981ne,Ahrens:1986xe} (orange dotted curve) and the recent ETM lattice QCD calculations~\cite{Alexandrou:2020okk,Alexandrou:2021wzv} (green dot-dot-dashed curve), where confidence bands at $95\%$ confidence level are also shown. Within error bands, all these experimental-data-based results (MiniBooNE, Dipole fit, BNL E734, and \emph{Quasi}-$\nu N$) are well consistent with each other, which in turn validates the correctness of Eq.~(\ref{Relations-NC-CC}). In contrast, the ETM lattice result of $G_A^Z(Q^2)$~\cite{Alexandrou:2020okk,Alexandrou:2021wzv} shows sizeable deviation from the other (i.e. MiniBooNE, Dipole fit, BNL E734 and \emph{Quasi}-$\nu N$) results.

To the best of our knowledge, there is currently no direct experimental data of the nucleon weak-neutral induced pseudoscalar FF $G_P^Z(Q^2)$ in the literature. To obtain $G_P^Z(Q^2)$, we need the knowledge of $G_P^W(Q^2)$ and $G_P^s(Q^2)$ according to Eq.~(\ref{Relations-NC-CC}). The result of $G_P^s(Q^2)$ is given in Eq.~(\ref{NC-Z-FF-s-quark}). The remaining task is to obtain $G_P^W(Q^2)$. In chiral perturbation theory ($\chi$PT), the nucleon weak-charged induced pseudoscalar FF $G_P^W(Q^2)$ from full chiral structure up to next-to-leading-order (NLO) is given by~\cite{Bernard:1994wn,Bernard:2001rs}
\begin{equation}
	\begin{aligned}\label{CC-ChPT-FF-GP}
		G_P^W(Q^2)
		&= g_{\pi^\pm pn} \frac{2(M_p+M_n) F_\pi }{Q^2+M_\pi^2} - 2G_A^W(0) \frac{(M_p+M_n)^2}{(M_A^W)^2 } + \mathcal O(Q^2;M_\pi^2),
	\end{aligned}
\end{equation}
where $g_{\pi^\pm pn}$ is the pseudoscalar pion-nucleon coupling constant, $M_p$ ($M_n$) is the proton (neutron) mass, $M_\pi $ is the charged pion mass, $F_\pi=f_{\pi}/\sqrt{2} \approx (92.0653 \pm 0.8485)~\text{MeV}$~\cite{ParticleDataGroup:2024cfk} is the pion decay constant for the $\pi^+ \to \mu^+ \nu_{\mu}$ reaction, and $M_A^W \approx (1.026 \pm 0.021)\,\text{GeV}$~\cite{Bernard:2001rs}. Based on the recent combined analysis of experimental data using chiral effective field theory for $f_{\pi^\pm pn}^2=0.0769(5)^a(0.9)^b$~\cite{Reinert:2020mcu}, we find~\cite{ParticleDataGroup:2024cfk}
\begin{equation}
	\begin{aligned}\label{gPiNN-constant}
		g_{\pi^\pm pn} = \frac{\sqrt{4\pi} (M_p + M_n) }{ M_\pi } f_{\pi^\pm pn} \approx (13.22613\pm 0.04369),
	\end{aligned}
\end{equation}
where uncertainties of $f_{\pi^\pm pn}^2$ from the first ($a$) and second ($b$) errors are added in quadrature.

Alternatively, we can also obtain $G_P^W(Q^2)$ by assuming the pion-pole dominance (PPD) hypothesis~\cite{Jang:2019vkm,Chen:2022odn}, which is based on the low-energy QCD relations---the partially conserved axial-vector current (PCAC) relation and the Goldberger-Treiman relation~\cite{Goldberger:1958tr}, namely
\begin{equation}
	\begin{aligned}\label{CC-PPD-FF-GP}
		G_P^W(Q^2)
		&= \frac{ (M_p+M_n)^2 }{ M_{\pi}^2 + Q^2 } G_A^W(Q^2),
	\end{aligned}
\end{equation}
where $G_A^W(Q^2)$ is extracted from quasielastic (anti)neutrino scattering data~\cite{Bernard:2001rs} via Eq.~(\ref{NC-Z-FF-GA}). We can thus construct $G_P^Z(Q^2) = [G_P^W(Q^2) - G_P^s(Q^2) ]/2$ with known $G_P^s(Q^2)$ (\ref{NC-Z-FF-s-quark}) by using $G_P^W(Q^2)$ either from Eq.~(\ref{CC-ChPT-FF-GP}) which is labeled by ``NLO $\chi$PT'', or from Eq.~(\ref{CC-PPD-FF-GP}) which is labeled by ``PPD + \emph{Quasi}-$\nu N$''. The PPD hypothesis also inspires us to reconstruct $G_P^Z(Q^2)$ by using directly the ``Dipole fit'' $G_A^Z(Q^2)$ from Eqs.~(\ref{NC-Z-FF-GA}) and (\ref{NC-Z-FF-MA}), namely
\begin{equation}
	\begin{aligned}\label{NC-PPD-FF-GP}
		G_P^Z(Q^2)
		&= \frac{ 4M^2 }{ M_{\pi}^2 + Q^2 } G_A^Z(Q^2),
	\end{aligned}
\end{equation}
which is labeled by ``PPD + Dipole fit''.

In the lower panel of Fig.~\ref{Fig_weakNeutral_GAGP}, we present our results of $G_P^Z(Q^2)$ by using different methods: ``PPD + Dipole fit'' (red dot-dashed curve), ``PPD + \emph{Quasi}-$\nu N$'' (blue solid curve), and ``NLO $\chi$PT'', in comparison with the reconstructed $G_P^Z(Q^2)=[G_P^W(Q^2)-G_P^s(Q^2)]/2$ by using $G_P^s(Q^2)$ (\ref{NC-Z-FF-s-quark}) and the experimental data of $G_P^W(Q^2)$ from the ordinary $\mu$-capture measurements~\cite{Bardin:1981cq,MuCap:2012lei} at\footnote{More rigorously, $Q^2 = \left[ \frac{(M_\mu + M_p)^2 - M_n^2 }{M_\mu (M_\mu+M_p)} - 1 \right] M_\mu^2 = \left[ \frac{ M_p (M_\mu + M_p)-M_n^2}{M_\mu(M_\mu + M_p ) } \right] M_\mu^2 \approx 0.88~M_\mu^2$, where $M_\mu$ is the muon mass.} $Q^2 \approx 0.88\,M_\mu^2$ in the $\mu^- + p \to n + \nu_\mu$ reaction labeled by ``$\mu$-capture (1981)''~\cite{Bardin:1981cq} (square marker) and ``$\mu$-capture (2013)''~\cite{MuCap:2012lei} (triangle marker), and from the low-energy charged pion electroproduction measurements~\cite{Choi:1993vt} labeled by ``Saclay exp.'' (circle markers). Besides, we also show the results of $G_P^Z(Q^2) = [G_P^{(u-d)}-G_P^s]/2$ from the recent ETM lattice QCD calculations~\cite{Alexandrou:2020okk,Alexandrou:2021wzv} labeled by ``ETM lattice'' (green dot-dot-dashed curve). We find that all these results of $G_P^Z$ are well consistent with each other, which also indicates the validity of Eq.~(\ref{Relations-NC-CC}). Owing to the intensive overlaps of these results, the confidence bands for $G_P^Z(Q^2)$ at $95\%$ confidence level are not shown in Fig.~\ref{Fig_weakNeutral_GAGP}.

For the nucleon weak-neutral induced pseudotensor FF $G_T^Z$, there is currently no direct experimental data at all. To some extent, this also explains why the second-class current contribution of the nucleon associated with the induced pseudotensor FF $G_T^Z(Q^2)$ are scarcely discussed and calculated in the literature~\cite{Meissner:1986xg,Meissner:1986js,Bernard:1994wn,Silva:2005fa,Schindler:2006jq,Aliev:2007qu,Sharma:2009hg,Eichmann:2011pv,Liu:2014owa,Dahiya:2014jfa,Ramalho:2015jem,Anikin:2016teg,Mamedov:2016ype,Hashamipour:2019pgy,Mondal:2019jdg,Zhang:2019iyx,Jun:2020lfx,Chen:2020wuq,Chen:2021guo,Ahmady:2021qed,Sauerwein:2021jxb,Xu:2021wwj,Atayev:2022omk,Chen:2022odn,Cheng:2022jxe,Liu:2022ekr,Ramalho:2024tdi}. According to the Fig.~7 of Ref.~\cite{Day:2012gb}, one can \emph{assume} that $G_T^W(Q^2)$ can by roughly approximated by $G_T^W(Q^2)\equiv \kappa_T G_A^W(Q^2)$, where the factor $\kappa_T \approx 0.1$ is roughly the mean value of the ratio $G_T^W(0)/G_A^W(0)$ in the Fig.~7 of Ref.~\cite{Day:2012gb} with $G_A^W = F_A$ and $G_T^W = 2F_A^{3}$, and $G_A^W(Q^2)$ is given in Eq.~(\ref{NC-Z-FF-GA}) with $M_A^W \approx (1.026 \pm 0.021)\,\text{GeV}$~\cite{Bernard:2001rs}. 

As a result, Ref.~\cite{Day:2012gb} thus inspires us to propose the following \emph{ans\"atz} for the nucleon weak-neutral induced pseudotensor FF $G_T^Z(Q^2)$:
\begin{equation}
	\begin{aligned}\label{NC-Z-FF-GT}
		G_T^Z(Q^2)
		&= \kappa_T G_A^Z(Q^2),
	\end{aligned}
\end{equation}
where $G_A^Z(Q^2)$ is the direct dipole model (\ref{NC-Z-FF-GA}) fit to the elastic (anti)neutrino-nucleon scattering data~\cite{MiniBooNE:2010xqw,MiniBooNE:2013dds,Sufian:2018qtw}. We should emphasize that the reason why we relate $G_T^Z(Q^2)$ to $G_A^Z(Q^2)$ via the \emph{ans\"atz} (\ref{NC-Z-FF-GT}) is simply because the assumption proposed in Ref.~\cite{Day:2012gb}, which is the \emph{only} reference that we have ever found with experimentally reasonable and useful relation for $G_T^W(Q^2)$ [and thus for $G_T^Z(Q^2)$]. In the real word, $G_T^Z(Q^2)$ is most probably to be quite different from $G_A^Z(Q^2)$ instead of the simple but naive scaling \emph{ans\"atz} (\ref{NC-Z-FF-GT}), since $G_T^Z(Q^2)$ is strongly constrained by many symmetries and conservation laws while $G_A^Z(Q^2)$ is not. This \emph{also} provides a key motivation for future experimental measurements of $G_T^Z(Q^2)$, e.g. using the formulas of full tree-level unpolarized differential cross sections~\cite{Chen:2024ksq}, for the proton in (anti)neutrino-proton elastic scattering.

For numerical calculations and illustrations of the proton weak-neutral axial-vector four-current distributions in Sec.~\ref{sec:Breit frame distributions} to Sec.~\ref{sec:Light-front distributions}, we declare that we employ only nucleon weak-neutral axial-vector FFs $G_A^Z(Q^2)$ from Eq.~(\ref{NC-Z-FF-GA}) using the ``Dipole fit'', $G_P^Z(Q^2)$ from Eq.~(\ref{NC-PPD-FF-GP}) using the ``Dipole fit'' $G_A^Z(Q^2)$ and the PPD hypothesis, and $G_T^Z(Q^2)$ from Eq.~(\ref{NC-Z-FF-GT}) using the ``Dipole fit'' $G_A^Z(Q^2)$ and $\kappa_T \approx 0.1$.

\section{Breakdown of Abel transformation for axial charge distributions}
\label{Appendix-B}

The Abel and its inverse transforms, named after Niels~H.~Abel for integral transforms in mathematics, have been visited recently~\cite{Moiseeva:2008qd,Panteleeva:2021iip,Kim:2021jjf,Kim:2021kum,Kim:2022bia,Kim:2022syw,Choudhary:2022den} in the case of charge and energy-momentum tensor spatial distributions with the goal of connecting 2D LF spatial distributions with the corresponding 3D ones, where 2D LF spatial distributions are regarded as the 2D Abel images of the corresponds 3D spatial distributions. We notice that some discussions and debates have been triggered in Refs.~\cite{Freese:2021mzg,Panteleeva:2021iip}. In this appendix, we will explicitly show the breakdown of Abel and its inverse transforms for the connection in physics between 2D LF and 3D BF axial charge distributions in the spin-$\frac{1}{2}$ case.

According to textbooks, the standard definitions of the Abel and its inverse transforms are given by~\cite{Bracewell2000bok}
\begin{equation}
	\begin{aligned}\label{AbelTransform-standard}
		\mathscr A\left[g \right] (b) 
		&\equiv \mathcal{G}(b) =2\int_{b}^{\infty}\ud r\frac{r}{\sqrt{r^2-b^2}}g(r),\\
		g(r) 
		&= -\frac{1}{\pi}\int_{r}^{\infty}\mathrm{d}b  \frac{1}{\sqrt{b^2-r^2}}\frac{\ud \mathcal{G}(b) }{\ud b},
	\end{aligned}
\end{equation}
where $\mathscr A\left[g \right] (b)\equiv \mathcal{G}(b)$ is called the 2D Abel image of the 3D spatial function $g(r)$~\cite{Freese:2021mzg}. It is thus not difficult to obtain the following generic relation for the $n$th order Mellin moment of the Abel image $\mathcal{G}(b)$ in connecting with the corresponding 3D spatial function $g(r)$:
\begin{equation}
	\begin{aligned}\label{Mellin-moments}
		\int_0^{\infty}\ud b\, b^{n-1}\, \mathcal{G}(b)
		=  \frac{\sqrt{\pi}\, \Gamma\left(\frac{n}{2} \right)}{\Gamma\left(\frac{n+1}{2} \right) } \int_0^{\infty} \ud r\, r^{n}g(r),\qquad n \in \mathbb N^+ =\{1,2,3,\cdots\},
	\end{aligned}
\end{equation}
which is believed to be valid as long as $g(r)$ decreases faster than any order of $r^{n}$~\cite{Kim:2021kum}. 

Using the dipole model \emph{ans\"atz} (\ref{NC-Z-FF-GA}) for the axial FF $G_A^Z(Q^2)$, one can easily obtain the following analytic expression of the LF axial charge distribution from Eq.~(\ref{2DLF-distributions}) for a longitudinally polarized spin-$\frac{1}{2}$ target (with $b=|\uvec b_\perp|$):
\begin{equation}
	\begin{aligned}\label{2DLF-Jp-Dipole-spinL}
		J_{5,\text{LF}}^{+}(\uvec b_\perp;P^+)
		&= (\sigma_z)_{\lambda' \lambda } G_A^Z(0) \frac{b (M_A^{Z})^3}{4\pi}\,  K_1(b M_A^{Z}) \equiv J_{5,\text{LF}}^{+}(b),
	\end{aligned}
\end{equation}
which is axially symmetric and can be regarded as the 2D Abel image of a 3D spatial distribution $J^{0}_{5,\text{naive}}(r)$. Applying the inverse Abel transform (\ref{AbelTransform-standard}) to $J_{5,\text{LF}}^{+}(b)$, we find
\begin{equation}
	\begin{aligned}\label{3DAbel-J0-spinL}
		J^{0}_{5,\text{naive}}(r)
		&= (\sigma_z)_{\lambda' \lambda } G_A^Z(0)\frac{ (M_A^{Z})^3}{8\pi}\, e^{-r M_A^{Z} }.
	\end{aligned}
\end{equation}

One can first check that the axial charge normalization condition
\begin{equation}
	\begin{aligned}
		(\sigma_z)_{\lambda' \lambda } G_A^Z(0) = \int\ud^3 r \, J^{0}_{5,\text{naive}}(r) =\int \ud^2 b_\perp \, J_{5,\text{LF}}^{+}(b)
	\end{aligned}
\end{equation}
seems to be automatically guaranteed. Besides, one can also check that $J_{5,\text{LF}}^{+}(b)$ and $J^{0}_{5,\text{naive}}(r)$ indeed satisfy the generic relation (\ref{Mellin-moments}) for the corresponding Mellin moment at any order. In particular, we find that the mean-square radius of $J^{0}_{5,\text{naive}}(r)$ is given by
\begin{equation}
	\begin{aligned}\label{Abel-Axial-J0-MSR}
		\langle r_A^2 \rangle_{\text{naive}}^\text{Abel}
		&= \frac{\int \ud^3r\, r^2 J^{0}_{5,\text{naive}}(r) }{ \int \ud^3r\, J^{0}_{5,\text{naive}}(r) } = \frac{12}{(M_A^{Z})^2} \approx (0.6510~\text{fm})^2\\
		&= - \frac{6}{G_A^Z(0)} \frac{\ud G_A^Z(Q^2)}{\ud Q^2} \bigg|_{Q^2=0} = R_A^2,
	\end{aligned}
\end{equation}
which exactly coincides with $R_A^2$~(\ref{naive-3DMS-axial-radius}) widely employed in the literature~\cite{Meissner:1986xg,Meissner:1986js,Bernard:1992ys,A1:1999kwj,Meyer:2016oeg,Hill:2017wgb,MINERvA:2023avz,Petti:2023abz,Kaiser:2024vbc}. In contrast to $J^{0}_{5,\text{naive}}$ (\ref{3DAbel-J0-spinL}), the \emph{genuine} 3D axial charge distribution (\ref{3DBF-distributions}) in the BF for a longitudinally polarized spin-$\frac{1}{2}$ target is in fact given by
\begin{equation}
	\begin{aligned}\label{3Dgenuine-J0-spinZ}
		J_{5,B}^0(\uvec r)
		&= (\sigma_z)_{s's}\int\frac{\ud^3\Delta}{(2\pi)^3}\, e^{-i\uvec\Delta \cdot \uvec r}\, \frac{ i\Delta_z }{2M} G_T^Z(\uvec\Delta^2),\\
	\end{aligned}
\end{equation}
which is actually related to the induced pseudotensor FF $G_T^Z(Q^2)$ rather than the axial FF $G_A^Z(Q^2)$. This explicitly demonstrates that even though we neglect the polarization difference, the naive 3D distribution $J^{0}_{5,\text{naive}}(r)$ does not assume clear physical meaning for quantifying the genuine 3D spatial distribution of weak-neutral axial charges in the BF for a longitudinally polarized spin-$\frac{1}{2}$ target. This explicitly reveals for the first time the breakdown of Abel and its inverse transforms for the connection in physics between 2D LF and 3D BF axial charge distributions in the spin-$\frac{1}{2}$ case. Our demonstration of the breakdown of Abel transformation is well consistent with the fact that the LF Galilean subgroup (singled out from the Lorentz group) does not have an $SO(3)$ subgroup, and thus LF dynamics does not preserve the 3D spherical symmetry~\cite{Brodsky:1997de,Freese:2021mzg}.

\acknowledgments

We warmly thank Dr.~Raza~Sabbir~Sufian for very helpful communications, and Profs. Dao-Neng Gao, Ren-You Zhang and Guang-Peng Zhang for valuable discussions at an early stage of this work. We are very grateful to Profs.~C\'edric~Lorc\'e, Qun Wang and Yang~Li for so many valuable encouragements, discussions and suggestions during our collaboration~\cite{Chen:2024oxx}. We thank Profs. Chueng-Ryong~Ji, Carlos~Mu\~{n}oz~Camacho, Weizhi~Xiong, Qin-Tao~Song, Tobias~Frederico, Jo\~{a}o~P.~de~Melo, Stanislaw~D.~G{\l}azek, Craig~D.~Roberts, Wayne~N.~Polyzou, Ismail~Zahed, Sanjin~Benic, Feng-Kun~Guo, Shan~Cheng, Yu~Jia and Drs.~Ho-Yeon Won, Poonam~Choudhary, Sudeep~Saha, Jani~Penttala for the helpful communications during the ``Light-Cone 2024: Hadron Physics in the EIC era'' conference. This work is supported in part by the National Natural Science Foundation of China (NSFC) under Grant Nos.~12135011, 11890713 (a sub-Grant of 11890710), and by the Strategic Priority Research Program of the Chinese Academy of Sciences (CAS) under Grant No.~XDB34030102.

\paragraph{Note added.} Recently, we were informed of Ref.~\cite{Panteleeva:2024vdw} which works on a similar topic.

%
\bibliographystyle{JHEP}
\bibliography{main_Refs.bib}

\end{document}